\documentclass[manuscript]{aastex}
\usepackage{lscape}

\slugcomment{Not to appear in Nonlearned J., 45.}

\shorttitle{Collapsed Cores in Globular Clusters}
\shortauthors{Djorgovski et al.}

\begin{document}


\title{Survey Observations of a Possible Glycine Precursor,  Methanimine (CH$_2$NH) }


\author{Taiki SUZUKI\altaffilmark{1}, Masatoshi OHISHI\altaffilmark{1,2},Tomoya HIROTA\altaffilmark{1,2}, Masao SAITO\altaffilmark{1,2}, Liton MAJUMDAR\altaffilmark{3,4,5}, Valentine WAKELAM\altaffilmark{3,4}}
\affil{$^1$ Department of Astronomy, the Graduate University for Advanced Studies (SOKENDAI), Osawa 2-21-1, Mitaka, Tokyo 181-8588, Japan}
\affil{$^2$ National Astronomical Observatory of Japan, Osawa 2-21-1, Mitaka, Tokyo 181-8588, Japan}
\affil{$^3$ Univ. Bordeaux, LAB, UMR 5804, F-33270, Floirac, France}
\affil{$^4$ CNRS, LAB, UMR 5804, F-33270, Floirac, France}
\affil{$^5$ Indian Centre for Space Physics, 43 Chalantika, Garia Station Road, Kolkata, 700084, India}

\email{taiki.suzuki@nao.ac.jp}



\begin{abstract}
We conducted survey observations of a glycine precursor, methanimine or methylenimine (CH$_2$NH), with the NRO 45~m telescope and the SMT telescope towards 12 high-mass and two low-mass star-forming regions  in order to increase number of CH$_2$NH sources and to better understand  the characteristics of CH$_2$NH sources.
As a result of our survey, CH$_2$NH was detected in eight sources, including four new sources.
The estimated fractional abundances were $\sim$10$^{-8}$ in Orion~KL and G10.47+0.03, while they were $\sim$10$^{-9}$ towards the other sources.
Our hydrogen recombination line and past studies suggest that CH$_2$NH-rich sources have less evolved HI\hspace*{-1pt}I regions.  
The less destruction rates by UV flux from the central star would be contributed to the high CH$_2$NH abundances towards CH$_2$NH-rich sources.
Our gas-grain chemical simulations suggest that CH$_2$NH is mostly formed in the gas-phase by neutral-neutral reactions rather than the product of thermal evaporation from the dust surfaces. 

\end{abstract}





\section{Introduction}
It is widely thought that prebiotic chemical evolution from small to large and complex molecules would have resulted in the origin of life.
There are two conflicting views regarding how complex organic molecules (COMs) were formed on the early Earth: formation on planetary surface or exogenous delivery.

\cite{Miller53} conducted the very famous experiment - Miller-Urey's experiment.
He demonstrated that organic materials can be synthesized from ``inorganic material", which was against the ``common sense" at that time.
He synthesized several organic compounds such as the amino acids glycine or alanine, by discharging a gas mixture of H$_2$O, NH$_3$, CH$_4$, and H$_2$, which was thought as the atmospheric composition of the early Earth.
However, recent studies have suggested that the amount of synthesized organic molecules would decrease dramatically under the currently suggested CO$_2$ dominant atmosphere \citep{Ehrenfreund02}.

In the interstellar medium (ISM), more than 190 molecules ranging from simple linear molecules to COMs were detected mainly towards dark clouds, low-mass and high-mass star-forming regions (CDMS: https://www.astro.uni-koeln.de/cdms/molecules).
\cite{Ehrenfreund02} argued that the exogenous delivery of COMs to the early Earth by comets and/or asteroids could be more than their terrestrial formation by three orders of magnitude; molecules delivered from the Universe might have played a crucial role in early Earth chemistry.
If this is the case, understanding of the interstellar chemistry will enable us to better understand the first stage of chemical evolution regarding to the origin of life: from atoms to very simple prebiotic species.

Among prebiotic species, great attention has been paid to amino acids, essential building blocks of terrestrial life; many surveys were made unsuccessfully to search for the simplest amino acid, glycine (NH$_2$CH$_2$COOH), towards Sgr B2 and other high-mass star forming regions \cite[e.g.,][and References therein]{Brown79,Snyder83,Combes96}.
Although \cite{Kuan03} claimed the first detection of glycine in 2003, several follow-up observations denied the detection \citep[e.g.,][]{ Jones07}.
The difficulty of the past glycine surveys would be due to potential weakness of glycine lines and the low sensitivities of telescopes used for the surveys.
In these days, the Atacama Large Millimeter/submillimeter Array (ALMA) is expected to overcome such difficulties associated with glycine surveys.
\cite{Garrod13} used his three phase chemical model that simulates fully coupled gas phase, grain surface and bulk ice chemistry and suggested the possibility in detecting glycine by using ALMA.

Formation processes of glycine have been studied by many researchers.
Theoretical and laboratory studies have demonstrated that glycine is formed on icy grain surface from methylamine (CH$_3$NH$_2$) and CO$_2$ under UV irradiation \citep{Holtom05}. 
It is suggested that CH$_3$NH$_2$ can be formed from abundant species, CH$_4$ and NH$_3$, on icy dust surface \citep{Kim11}.
Alternatively methanimine (CH$_2$NH) would be related to the formation of CH$_3$NH$_2$.
Another possible route to form these species is hydrogenation to HCN on the dust surface, as investigated both in experimentally and theoretically \citep{Woon02, Theule11}.
These issues also have been discussed in details by \cite{Majumdar13} where they identified  CH$_2$NH$_2$ as another possible precursor for glycine formation.

HCN $\stackrel{\rm 2H}{\longrightarrow}$ CH$_2$NH

CH$_2$NH $\stackrel{\rm 2H}{\longrightarrow}$ CH$_3$NH$_2$

\cite{Theule11} mentioned that the hydrogenation process to CH$_2$NH might be much quicker compared to that of HCN.
It is also suggested that CH$_2$NH may be formed in the gas phase reactions among radicals \citep{Turner99, Halfen13}.
\cite{Halfen13} showed that the excitation temperatures of CH$_2$NH and CH$_3$NH$_2$ are different, suggesting different formation paths for CH$_2$NH and CH$_3$NH$_2$.
If this is the case, CH$_2$NH might not be a direct precursor to CH$_3$NH$_2$.

\begin{figure}
\includegraphics[scale=.6]{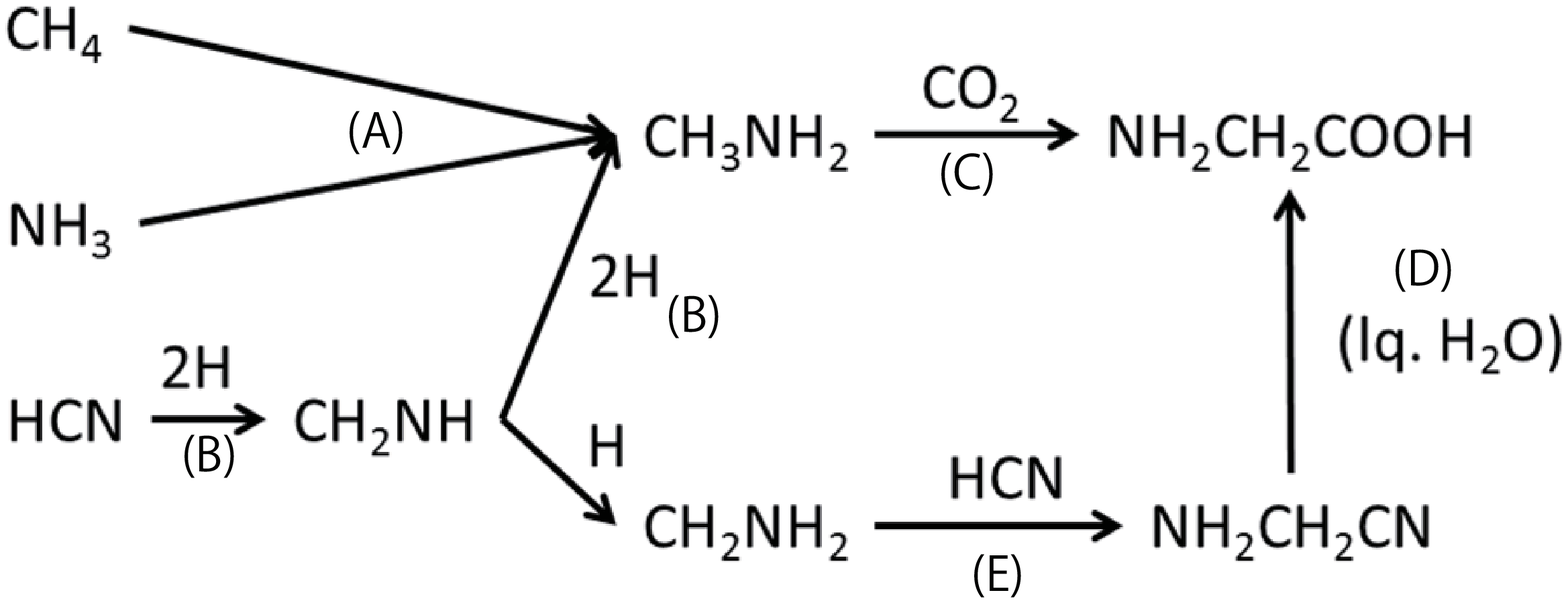}
\caption{
Assumed formation scenario to glycine.
``lq" represents the hydrolysis process.
References: (A)\cite{Kim11} (B)\cite{Theule11} and \cite{Woon02} (C)\cite{Holtom05} (D)\cite{Peltzer84} (E) \cite{Danger11}
\label{fig:formation}
}
\end{figure}

CH$_2$NH is related to another formation path to glycine;
if H atom and HCN are added to CH$_2$NH on the grain surface, aminoacetonitrile (NH$_2$CH$_2$CN) would be formed \citep{Danger11}.
NH$_2$CH$_2$CN yields glycine via hydrolysis \citep{Peltzer84}.
These formation processes are shown in Figure~\ref{fig:formation}.

Such precursor candidate molecules were reported in the ISM.
First detection of CH$_2$NH was made towards Sgr B2 \citep{Godfrey73}.
CH$_2$NH was confirmed towards Sgr B2 by many authors, and in particular, \cite{Halfen13} reported 27 transitions of CH$_2$NH towards Sgr B2.
\cite{Dickens97} reported new detections of CH$_2$NH in three high-mass star forming regions, W51~e1/e2, Orion KL, and G34.3+0.15.
Recently, \cite{Qin10} reported the detection of CH$_2$NH towards G19.61-0.23 through a line survey with the Sub-Millimeter Array.
The past results of CH$_2$NH surveys are summarized in Table~\ref{past_CH2NHlines}.
Further CH$_3$NH$_2$ was detected towards Sgr B2 \citep{Kaifu74} and Orion KL \citep{Fourikis74}.
\cite{Isokoski13} conducted a line survey towards G31.41+0.3, and reported detection of CH$_3$NH$_2$. 
NH$_2$CH$_2$CN was detected towards Sgr B2 \citep{Belloche08}.

Investigation of glycine formation processes described in Figure~\ref{fig:formation} would be an important step to know how ``seed of life" is formed from ubiquitous species. 
One of the ways to discuss the formation processes would be to compare actual CH$_2$NH abundances in various physical conditions with the chemical modelings.
However, the small number of CH$_2$NH sources makes it difficult to discuss the connections of different physical conditions and CH$_2$NH chemistry.
The survey observations of possible glycine precursors will help not only to discuss the physical condition where CH$_2$NH is rich, but also to plan promising future glycine surveys with ALMA.

In this paper, we extended the CH$_2$NH survey conducted by \cite{Dickens97} towards high-mass and low-mass star forming regions, and report four new CH$_2$NH sources.
Our new detection of CH$_2$NH sources enabled us to discuss the connection of CH$_2$NH chemistry and their physical properties. 
The observational method will be described in section 2 (Observations); our results will be described in section 3 (Results); we will compare the evolutionary stages of the CH$_2$NH sources in section 4 (Discussion).
We will also discuss the formation path to CH$_2$NH in the discussion section.
We will summarize our work in section 5 (Conclusion).

\section{Observations}
\subsection{Observation Procedure}
We conducted survey observations of CH$_2$NH with the 45 m radio  telescope at the Nobeyama Radio Observatory (NRO), National Astronomical Observatory of Japan in April, May, and December, 2013, and the Sub-Millimeter Radio Telescope (SMT), Arizona Radio Observatory in October, 2013.
The observed lines and their parameters are summarized in Table~\ref{CH2NHlines-summary}.
Various species are simultaneously observed in our observations.
The observations and analysis of other species will be presented and compared with the chemical modeling in a separate paper, to discuss chemistry to form glycine.

At NRO, we used the TZ receiver \citep{Nakajima13} in the dual polarization mode.
The 4$_{ 0 4}$-3$_{ 1 3}$ transition at 105~GHz was observed.
The SAM45 spectrometer was used for the backend with a frequency resolution of 122~kHz, corresponding to the velocity resolution of 0.35~km~s$^{-1}$ at 105~GHz.
The system temperature ranged from 150 to 300~K.
The main beam efficiency ($\eta_{\rm mb}$) was 0.31, and the beam size (FWHM) was 16".
Observations were performed in the position switching mode.
The pointing accuracy was checked by observing nearby SiO masers, and the pointing error was typically 5", but 10" under windy condition.

With the SMT telescope, the 1$_{ 1 1}$-0$_{ 0 0}$, 4$_{ 1 4}$-3$_{ 1 3}$, and 4$_{ 2 3}$-3$_{ 2 2}$ lines at 225, 245, and 255~GHz regions, respectively, were observed.
Their parameters are summarized in Table~\ref{CH2NHlines-summary}.
We used the 1 mm ALMA SBS SIS receiver in the dual polarization mode.
The system temperature was between 250 and 400~K.
The main beam efficiency was 0.74. 
The beam sizes (FWHM) were 35" and 40" depending on the observing frequencies.
The position switching mode was employed.
The pointing accuracy was checked using the Venus, and its accuracy was kept within 10".
We used the Forbes Filters as the spectrometer with a frequency resolution of 250~kHz, which is equivalent to the velocity resolution of 0.29 km~s$^{-1}$.
Since the Forbes Filter has 512 channels, the observed frequency range covered 128~MHz wide at 225~GHz.
For the transitions in 245 and 255~GHz, we split the Forbes filter into two 256 channel sets, and two lines were observed simultaneously.

\subsection{Source Selection}
\cite{Theule11} suggested that CH$_2$NH is formed via the hydrogenation process to HCN (HCN $\rightarrow$ CH$_2$NH).
To select our target sources, we noted that CH$_3$OH is well known to be formed via hydrogenation process to CO on the grain surface (CO $\rightarrow$ H$_2$CO $\rightarrow$ CH$_3$OH).
If the formation process to CH$_2$NH is similar to that of CH$_3$OH, CH$_2$NH would be potentially rich towards CH$_3$OH-rich sources.
Thus, our sources were selected from ``The Revised Version of Class I Methanol Maser Catalog" \citep{Valtts12}, which is available on the Internet: (http://www.asc.rssi.ru/mmI/), and from CH$_3$OH maser sources \citep{Minier02}.
We also took into account the results of survey observations of complex organic molecules towards hot core sources by \cite{Ikeda01}. 
Hot cores are high-mass star forming regions traced by the high temperature and dense molecular gas, which are sometimes associated with HI\hspace*{-1pt}I regions.
Further, we also selected two famous low-mass star forming regions where complex molecules have been observed.
These 14 sources for our survey are summarized in Table~\ref{sources}.

In this table, W3(H$_2$O), Orion~KL, NGC6334F, G10.47+0.03, G19.61-0.23, G31.41+0.3, W51~e1/e2, and DR21(OH) are molecule-rich sources \citep{Ikeda01}.
G75.78+0.34 and Cep~A were selected from the CH$_3$OH maser catalogs.
The above sources are high-mass star-forming regions.

\section{Results}
\begin{figure}
\includegraphics[scale=.35]{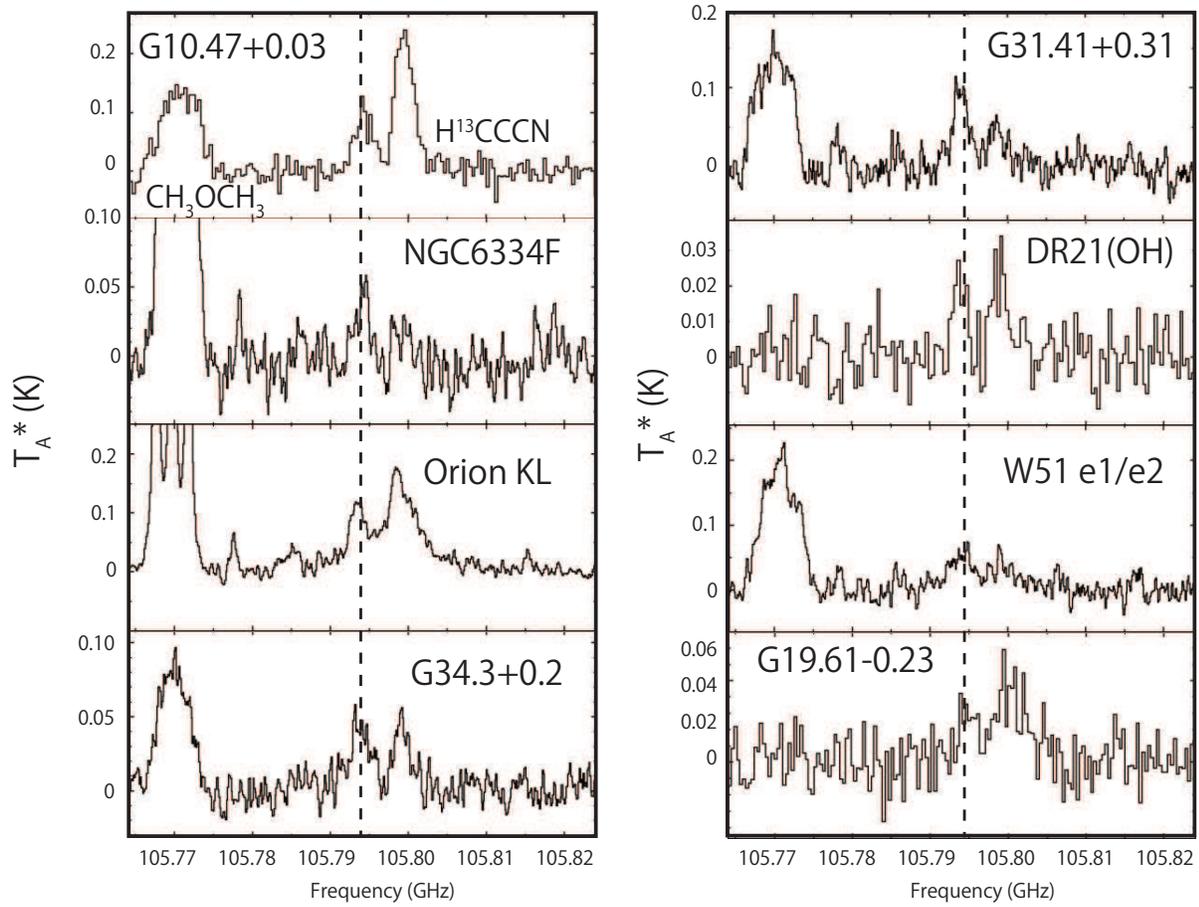}
\caption{
Observed CH$_2$NH 4$_{0 4}$-3$_{ 1 3}$ transitions towards the detected sources.
The vertical lines represent the rest frequency, 105.794062~GHz.
\label{fig:CH2NHspectral}
}
\end{figure}

We detected CH$_2$NH lines towards eight high mass star-forming regions as shown in Figure~\ref{fig:CH2NHspectral}.
Among them, Orion KL, W51~e1/e2, G34.3+0.15 and G19.61-0.23 are known CH$_2$NH sources in the past surveys \citep{Dickens97,Qin10}. 
However, we report the detection of the 4$_{ 0 4}$-3$_{ 1 3}$ transition for the first time.
Other sources, NGC6334F, G10.47+0.03, G31.41+0.3, and DR21(OH), are new CH$_2$NH sources.
The observing results are summarized in Table~\ref{CH2NHline-obs}.

\begin{figure}
\plotone{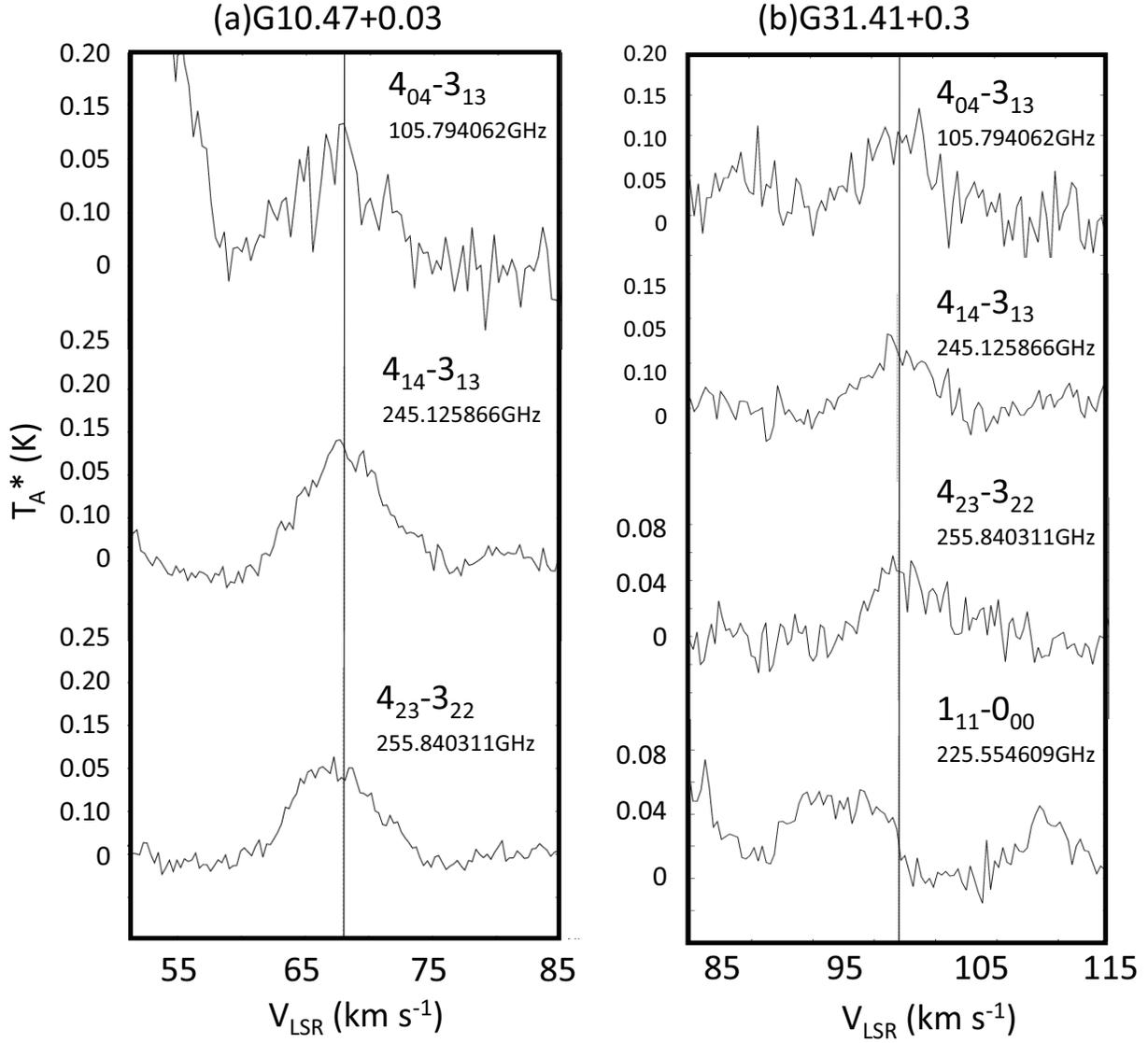}
\caption{
Comparison of radial velocities of CH$_2$NH lines observed with the NRO 45m telescope and the SMT telescope.
The vertical solid lines represent the systematic radial velocities of individual sources, shown in Table~\ref{sources}.
\label{fig:velocity-compare}
}
\end{figure}

Figure~\ref{fig:CH2NHspectral} shows detected 4$_{0 4}$-3$_{ 1 3}$ line of CH$_2$NH with the NRO 45~m telescope towards the eight high-mass star-forming regions.
The dotted lines correspond to the rest frequency of the 4$_{ 0 4}$-3$_{ 1 3}$ transition (105.794062~GHz).
Two other CH$_2$NH lines (4$_{ 1 4}$-3$_{ 1 3}$ and 4$_{ 2 3}$-3$_{ 2 2}$) were detected with the SMT telescope towards G10.47+0.03 and G31.41+0.3.
We note that the 4$_{0 4}$-3$_{ 1 3}$ line towards G10.47+0.03 and G31.41+0.3 show comparable intensities to that of Orion KL.
Figure~\ref{fig:velocity-compare} shows the CH$_2$NH lines observed towards G10.47+0.03 and G31.41+0.3 with the NRO 45~m telescope and the SMT telescope.
In this figure, the vertical solid line corresponds to the systematic radial velocity of each source, shown in Table~\ref{sources}.
The differences of radial velocities among the three observed transitions, $4_{0 4}$-$3_{1 3}$, $4_{1 4}$-$3_{1 3}$, and $4_{2 3}$-$3_{2 2}$, are within 0.25~km~s$^{-1}$.
On the other hand, we did not detect the $1_{1 1}$-$0_{0 0}$ transition towards G31.41+0.3.

In the other sources  (Cep A, W3(H$_2$O), W3(OH), G75.78+0.34, NGC2264 MM3 IRS1, NGC1333 IRAS4B, and IRAS16293-2422), no CH$_2$NH transitions were detected.
We briefly summarize their properties here.
Cep~A is a high mass star-forming region which harbors several radio continuum sources including several HI\hspace*{-1pt}I regions.
These complex components are spatially unresolved in our beam.
W3(H$_2$O) is an H$_2$O maser source with which HI\hspace*{-1pt}I regions are associated.
G75.78+0.34 is known to contain at least two ultra compact (UC) HI\hspace*{-1pt}I regions ionized by B0-B0.5 ZAMS stars and possibly Hyper-Compact HI\hspace*{-1pt}I regions.
NGC2264 MMS3 IRS1 is a dense core associated with a small star cluster, where HCOOCH$_3$ emission was reported \citep{Sakai07}.
NGC1333 IRAS4B and IRAS16293-2422 are class 0 low-mass protostars where complex organic molecules have been reported.
It is known that IRAS16293-2422 has three components named A1, A2, and B, which are unresolved  in our beam.
Outflow activities are confirmed in A2 and B \citep{Loinard13}.
Complex organic molecules such as methyl formate (HCOOCH$_3$) and glycolaldehyde (CH$_2$OHCHO) were detected towards IRAS16293-2422 \citep{Jorgensen12}.
NGC1333 IRAS4B is located 30" away from IRAS4A, enabling us to separate them with our beam.
HCOOCH$_3$ is detected towards NGC1333 IRAS4B \citep{Sakai06}.

\subsection{Derivation of Abundance}
In this subsection we will describe the methodologies in deriving fractional abundances of CH$_2$NH.
In the following discussion, we will show that reliable CH$_2$NH abundances can be obtained when we employ the same source size as that of continuum emission.
If continuum data are not available, we assume a source size of 10" in the derivation.
The estimated CH$_2$NH abundances under compact source sizes, which will be described later in detail, and a source sizes of 10" are summarized in Tables~\ref{CH2NH-abundance-compact} and \ref{CH2NH-abundance-10sec}, respectively.

In sources where more than two transitions are available, we calculated column densities of CH$_2$NH using the rotation-diagram method described in \cite{Turner91}, and the following equation was employed:
\begin{equation}
\log \frac{3kW}{8\pi^3 \nu S\mu^2 g_{\rm I}g_{\rm K}} = \log \frac{N}{Q_{\rm rot}} - \frac{E_{\rm u}}{k} \frac{\log e}{T_{\rm rot}}
\end{equation}
where $W$ is the integrated intensity, $S$ is the intrinsic line strength, $\mu$ is the permanent electric dipole moment, $g_{\rm I}$ and $g_{\rm K}$ are the nuclear spin degeneracy and the $K$-level degeneracy, respectively, $N$ is the column density, $Q_{\rm rot}$ is the rotational partition function, $E_{\rm u}$ is the upper level energy, and $T_{\rm rot}$ is the rotation temperature.
$g_{\rm I}$ and $g_{\rm K}$ are unity for CH$_2$NH because CH$_2$NH is an asymmetric top with no identical H atoms.
The column density and the excitation temperature can be derived by utilizing the least-squares fitting.
The column density is derived from the interception of a diagram, and its slope will give us the excitation temperature.  

In the excitation analysis above, the beam coupling factors would be key parameters.
We investigated the distribution of the CH$_2$NH 1$_{1 1}$-0$_{ 0 0}$ transition at 225.55455~GHz towards Orion~KL using the calibrated Science Verification data of ALMA cycle 0 (https://almascience.eso.org/almadata/sciver/OrionKLBand6/) to estimate CH$_2$NH source sizes.
The data were imaged with the Common Astronomy Software Applications (CASA) package.
This data have the spectral resolution of 488~kHz and the synthesized beam size of $\sim$2".
We selected line-free channels for continuum subtraction in the (u, v) domain using the CASA task, uvcontsub.
The channel maps of CH$_2$NH 1$_{1 1}$-0$_{ 0 0}$ distribution were made by the CASA task, CLEAN.
We employed the equal weighting of uv, and set the threshold value to stop CLEAN task at 72 mJy considering the r.m.s level.
After that, we made the integrated intensity map from 225.55348 to 225.55641~GHz, corresponding to the V$_{\rm LSR}$ range from 6.5 to 10.4~km~s$^{-1}$.

\begin{figure}
\includegraphics[scale=.6]{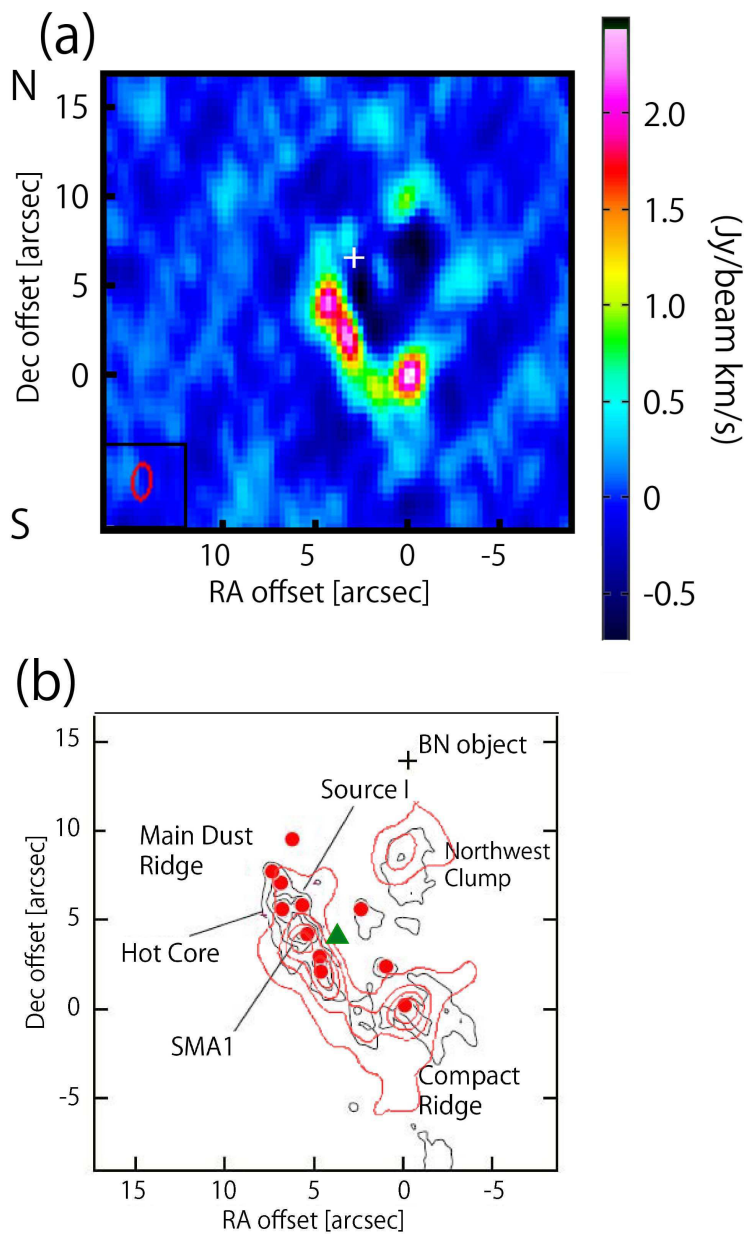}
\caption{
(a): The integrated intensity map of CH$_2$NH 1$_{ 1 1}$-0$_{ 0 0}$ transition. The velocity range is from 6.5 to 10.4 km~s$^{-1}$ and the spatial resolution is 2".  The (0, 0) position corresponds to R.A.(J2000) = 05$^{\rm h}$ 35$^{\rm m}$ 14$^{\rm s}$.1250, Dec.(J2000) = -05$\degr$ 22' 36".486. Our observing position by NRO 45~m telescope is shown by the cross. (b): The red contour map of CH$_2$NH $1_{1 1}$-$0_{0 0}$ transition is overlapped on the dust continuum emission obtained by \cite{Hirota15}. The (0, 0) position is the same as (a). The red circles in (b) correspond to the continuum sources reported by \cite{Hirota15}. A green triangle represents the position of infrared source called ``Source n". 
\label{fig:integrated_intensity_Orion}
}
\end{figure}

Our integrated intensity map of Orion~KL is shown in Figure~\ref{fig:integrated_intensity_Orion}(a).
The region corresponding to the beam of NRO 45~m telescope in our survey is shown by the black circle.
We found that there are three strong CH$_2$NH peaks in this region, which are unresolved with the synthesized beam of this data ($\sim$2").
In Figure~\ref{fig:integrated_intensity_Orion}(b), CH$_2$NH distributions are compared with the 0.8~mm dust continuum emission in Figure~4 of \cite{Hirota15}.
We found that the peak positions of CH$_2$NH are both in Orion Hot Core and Compact Ridge.
There is also a weak peak of CH$_2$NH in Northwest clump.
Since the distribution of CH$_2$NH and dust continuum emission look similar to each other, dust continuum emissions would be good indicators to estimate the source sizes of CH$_2$NH.
Thus, if dust continuum data were available, we employed the source size of dust continuum emission for CH$_2$NH to calculate the beam coupling factors.
In this case, column densities of molecular hydrogen were estimated from the dust continuum emission to calculate the fractional abundances of CH$_2$NH (Table~\ref{CH2NH-abundance-compact}).

\begin{figure}
\includegraphics[scale=.8]{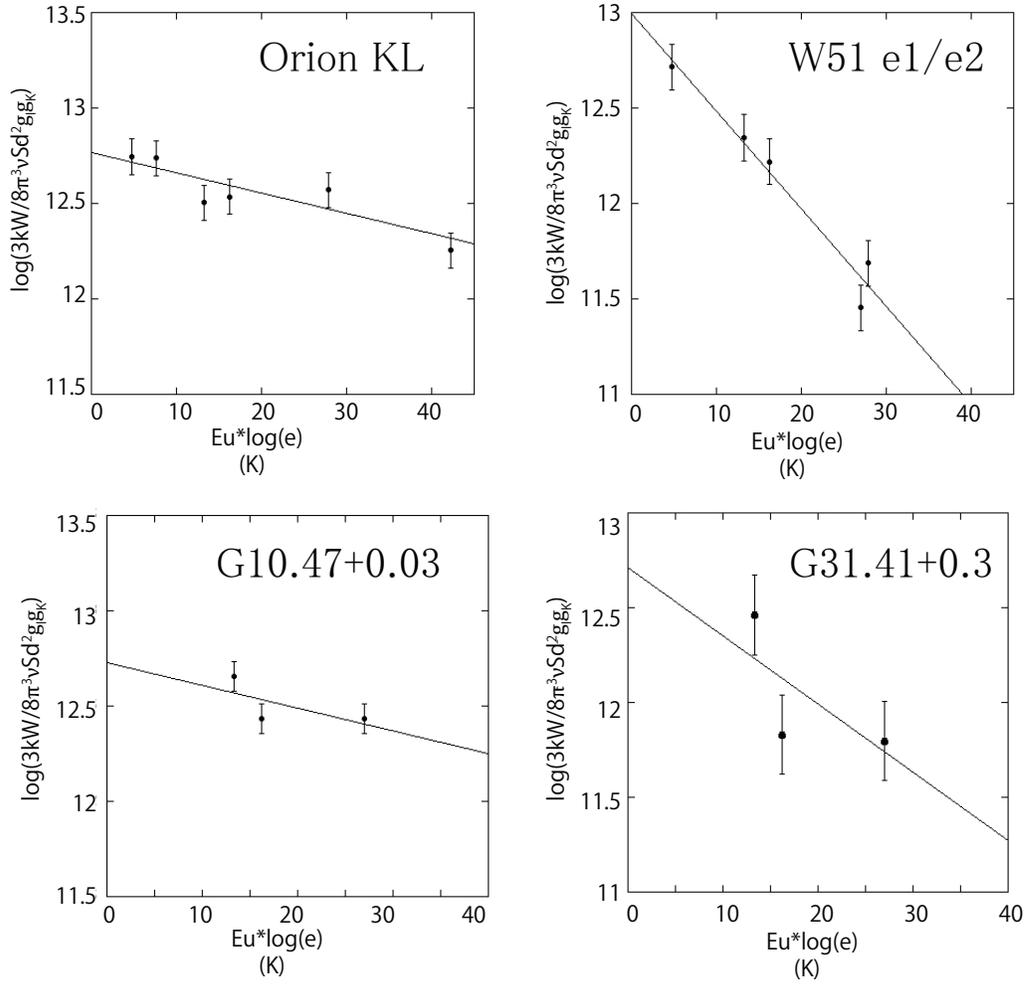}
\caption{
Rotation diagrams for Orion KL, W51~e1/e2, G10.47+0.03, and G31.41+0.3.
The plots are made using the data reported in \cite{Dickens97} in Orion~KL and W51~e1/e2.
\label{fig:rotation_diagrams}
}
\end{figure}

For G10.47+0.03 and G31.41+0.3, we made rotation diagrams to derive the column densities using the three transitions that we have observed.
On the other hand, for Orion KL and W51~e1/e2, we made rotation diagrams by including other CH$_2$NH transitions reported by \cite{Dickens97}.
These results are shown in Figure~\ref{fig:rotation_diagrams}.
For these four CH$_2$NH sources, we derived excitation temperatures ranging from 20~K to 90~K.
Since only one transition was observed towards the other sources, we assumed that the excitation temperature is similar to that of HCOOH (40~K) reported by \cite{Ikeda01} because the dipole moment of HCOOH is close to that of CH$_2$NH.
It would be important to know the error associated with the fractional abundances when the excitation temperature is fixed to be 40~K.
Considering that the wide range of excitation temperatures were derived from the rotation diagram method, we assumed that the excitation temperatures can be changed within $\pm$20~K, which can change the column densities by 40$\%$.
We note that the optical depths calculated with our method were less than 0.3 even for strong CH$_2$NH transitions.
The CH$_2$NH column densities and the fractional abundances derived assuming the compact sources are summarized in Table~\ref{CH2NH-abundance-compact}.

For sources where dust continuum emission data are not available (G34.3+0.2, DR21(OH), W3(H$_2$O), G75.78+0.34, Cep~A, NGC2264 MMS3, NGC1333 IRAS4B, and IRAS16293-2422), we assumed, based on extended CO data, that CH$_2$NH are spatially extended (source size of 10") in calculating the fractional abundances. 
In this case, both CH$_2$NH and molecular hydrogen may be underestimated due to beam dilution.
It would be important to know how fractional abundances tend to be affected if we employ the source size of 10".
For this purpose, we calculated CH$_2$NH fractional abundances towards Orion KL, G10.47+0.03, G31.41+0.3, NGC6334F, and W51~e1/e2 assuming source sizes of 10" in the upper side of Table~\ref{CH2NH-abundance-10sec}, and the CH$_2$NH fractional abundance ratio of compact source to 10" source were investigated. 
As a result, we found that the differences of the fractional abundances are within a factor of $\sim$3 (Table~\ref{CH2NH-abundance-10sec}).
Considering that the dependence on the source sizes is small, we employed a compact source size when dust continuum data are available (Table~\ref{CH2NH-abundance-compact}), and we used the source size of 10" to derive the CH$_2$NH fractional abundances for the other sources (the lower side of Table~\ref{CH2NH-abundance-10sec}).

\subsection{CH$_2$NH Abundances}
\begin{figure}
\includegraphics[scale=.6]{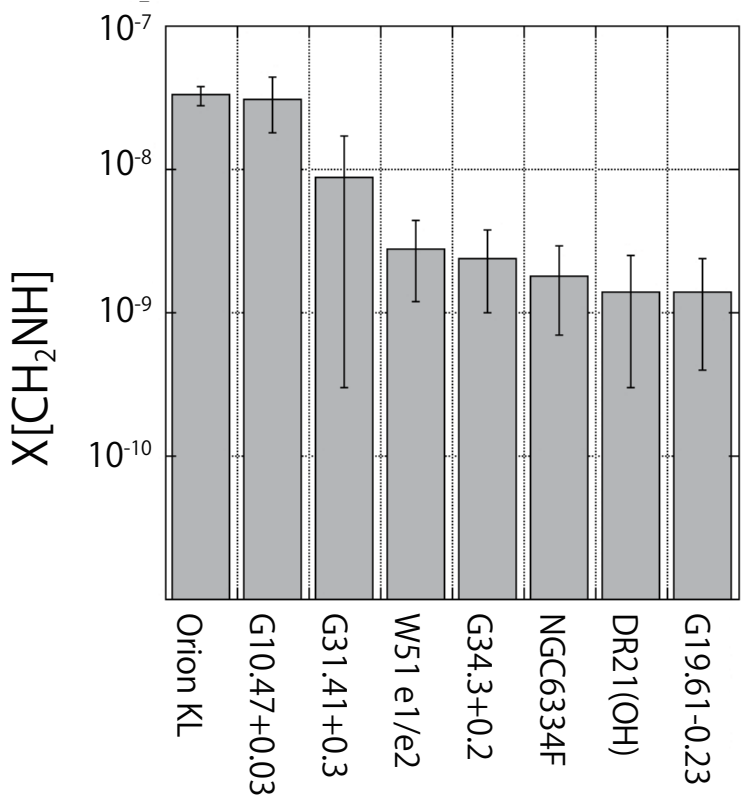}
\caption{
Histogram of CH$_2$NH fractional abundances.
The source size of 10" was assumed for G34.3+0.2 and DR21(OH).
For the other sources, compact source sizes were assumed.
The error bars represent the uncertainties associated with the derivation of CH$_2$NH abundances (i.e., r.m.s noise and the error associated with the scattering of the rotation diagrams, and assumption of the excitation temperatures.)
\label{fig:fractional_abundance}
}
\end{figure}

In Tables~\ref{CH2NH-abundance-compact} and ~\ref{CH2NH-abundance-10sec}, we summarize the results of our analysis.
Fractional abundances of CH$_2$NH are compared in Figure~\ref{fig:fractional_abundance}.
Our CH$_2$NH fractional abundance is the highest in Orion~KL.
The column density of CH$_2$NH towards Orion~KL is 6.0 ($\pm$1.8) $\times$ 10$^{14}$ cm$^{-2}$ in \cite{Dickens97} , and our value assuming a beam filling factor of unity is 1.2 ($\pm$0.4) $\times$ 10$^{15}$ cm$^{-2}$, which agrees within a factor of 2.
In our calculation under compact source size, the column density and the corresponding fractional abundance are, respectively, 5.5 ($\pm$0.8) $\times$ 10$^{15}$ cm$^{-2}$ and 3.3 ($\pm$0.5) $\times$ 10$^{-8}$.

G10.47+0.03 and G31.41+0.3 are CH$_2$NH-rich sources next to Orion~KL.
In G10.47+0.03, the fractional abundance is 3.1 ($\pm$1.1) $\times$ 10$^{-8}$.
G31.41+0.3 has the fractional abundance of 8.8 ($\pm$8.5) $\times$ 10$^{-9}$.
The large error of the fractional abundance in G31.41+0.3 is due to scattering of the plots in its rotation diagram (Figure~\ref{fig:rotation_diagrams}).
Transitions in much higher energy levels will enable us to determine the column densities more accurately.

The fractional abundances of other sources are, 2.4 ($\pm$1.6) $\times$ 10$^{-9}$, 2.8 ($\pm$1.6) $\times$ 10$^{-9}$, 2.4 ($\pm$1.7) $\times$ 10$^{-9}$, 1.4 ($\pm$1.1) $\times$ 10$^{-10}$, and 1.4 ($\pm$1.0) $\times$ 10$^{-9}$, in G34.3+0.2, W51~e1/e2, NGC6334F,  DR21~(OH), and G19.61-0.23, respectively.
If an extended source size was assumed, our column density towards W51~e1/e2, 6.6 ($\pm$3.6) $\times$ 10$^{13}$ cm$^{-2}$, is close to that reported by \cite{Dickens97}, 8 ($\pm$0.4) $\times$10$^{13}$ cm$^{-2}$.
From these results, we found that CH$_2$NH abundances range from $\sim$10$^{-9}$ to $\sim$10$^{-8}$.

CH$_2$NH was not detected with sufficient S/N ratios towards other sources (W3(H$_2$O), G75.78+0.34, Cep~A, NGC2264 MMS3, NGC1333 IRAS4B, IRAS16293-2422).
We derived upper limits to the column densities of CH$_2$NH for these sources corresponding to the three sigma r.m.s noise levels and excitation temperatures of 40~K.
We summarized these results in Table~\ref{CH2NH-abundance-10sec}.

\section{Discussion}
\subsection{Characteristics of CH$_2$NH-rich Sources}
As presented in Figure~\ref{fig:fractional_abundance}, the observed CH$_2$NH abundances are different by a factor of about 20.
What characteristics cause such a large difference among CH$_2$NH abundances?
In this section, we will discuss three hypotheses to explain the different CH$_2$NH abundances in terms of distance, kinetic temperatures, and evolutionary phase of sources.

\begin{figure}
\includegraphics[scale=.6]{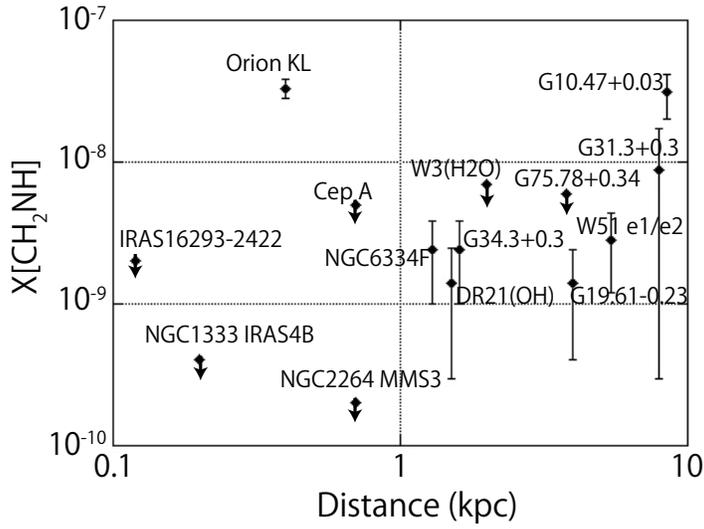}
\caption{
Plots of the distances to the sources against the CH$_2$NH fractional abundances.
The upper limits of X[CH$_2$NH] are shown for sources where CH$_2$NH is not detected.
\label{fig:distance}
}
\end{figure}

First of all, we will discuss the possibility that the differences of CH$_2$NH abundances are simply originating from the different distances to the sources.
Actually, the beam averaged CH$_2$NH column densities would be high for closer sources since they are less affected by beam dilution.
However, the differences depending on the source sizes were small when we compared their CH$_2$NH fractional abundances.
In Figure~\ref{fig:distance}, we plotted the CH$_2$NH fractional abundances against the distance to the sources.
It would be difficult to find the correlation between CH$_2$NH fractional abundances and the distances in Figure~\ref{fig:distance}.
The difference of CH$_2$NH fractional abundances would be due to source properties rather than just an effect of beam dilution.

The second possibility is that the different temperatures of CH$_2$NH sources might contribute to CH$_2$NH abundances.
\cite{Theule11} demonstrated that CH$_2$NH can be formed via the hydrogenation process to HCN on the dust surface (Figure~\ref{fig:formation}).
If CH$_2$NH is produced on dust surface, gas phase CH$_2$NH abundance would be lower until source temperature gets high enough to sublimate species from dust surface.
\cite{Hernandez14} reported rotation temperatures of CH$_3$CN, which is often used as a tracer of the kinetic temperature, towards 17 sources including G10.47+0.03, NGC6334F, W51~e2, and W3(H$_2$O).
The rotation temperatures were 499~K for G10.47+0.03, 408~K for NGC6334F, 314~K for W51~e2, and 182~K for W3(H$_2$O).
It is notable that W3(H$_2$O), where CH$_2$NH was not detected, has a lower temperature than CH$_2$NH sources.
Thus, this scenario would be convincing when the desorption temperature of CH$_2$NH is high ($>$180~K).
One easy way to roughly estimate desorption temperature would be to use the Clausius-Clapeyron equation (2) and the ideal gas law (3), described below:
\begin{equation}
\log_{10} {\rm P} = 4.559(1-({\rm T}/{\rm T_B}))
\end{equation}
\begin{equation}
{\rm P} = 8.314{\rm n}{\rm T}
\end{equation}
where T$_{\rm B}$ is the boiling point under the standard state, T is the sublimation temperature under the interstellar gas pressure P, and n is the number density of the species.
We have to be cautious that the desorption temperature derived in this method is an approximated value, and it would have an uncertainty in two reasons.
First, the Clausius-Clapeyron equation is applicable for the phase transition between pure solid or liquid and gas.
However, the actual interstellar molecules would be far from pure material, which are physisorbed on H$_2$O ice.
In spite of the difference of the physical condition, the Clausius-Clapeyron equation succeeds to reproduce the commonly accepted sublimation temperature of H$_2$O (from 90 to 100~K) and CO (from 15 to 18~K) in the ISM, assuming the density from 1$\times$10$^5$ to 1$\times$10$^7$ cm$^{-3}$.
Second, since the boiling point for CH$_2$NH is not measured, we had to use theoretically predicted value of 215($\pm$15)~K from a chemical database (http://www.chemspider.com/).

If we use the CH$_2$NH fractional abundance of 3.1$\times$10$^{-8}$ (the value observed in G10.47+0.03) and the molecular hydrogen number density of 10$^{7}$ cm$^{-3}$, the CH$_2$NH number density n is 0.31~cm$^{-3}$.
In our calculation, we assume the initial T values of from 20~K to 200~K (=T$_0$) and the P is obtained from the equation (2), then, next desorption temperature, T$_1$ is obtained with the equation (3).
In these iterations between equations (2) and (3), we are able to determine the final T value when the difference of successive temperatures reaches within 1~K.
As a result, we found that the estimated desorption temperature of 40~K is quite low compared to the gas temperatures of star-forming regions.

We also calculated the desorption temperature in terms of the equilibrium of microscopic accretion and desorption process, which may better approximate the interstellar situation.
The accretion rate R$_{\rm acc}$(i) (cm$^{-3}$ $s^{-1}$) and the desorption time scale t$_{\rm evap}$ (s) for the species $i$ are described in \cite{Hasegawa92} as shown below:
\begin{equation}
R_{\rm acc}(i) = \sigma_d <v(i)>n_{\rm gas}(i) n_d
\end{equation}
\begin{equation}
t_{\rm evap} = \nu_0^{-1}\exp(E_D/kT_{\rm dust}),
\end{equation}
\begin{equation}
{\rm where}~\nu_0 = (2n_sE_D/\pi^2m).
\end{equation}
$\sigma_d$, $<$v(i)$>$, n$_{\rm gas}$(i), n$_{\rm dust}$(i), n$_d$, E$_D$, n$_s$, m, respectively, represent the cross section of the dust, averaged velocity of the gas species under a certain temperature, the gas and the dust surface phase number density of the species, the dust number density, the desorption energy, the surface density of the site ($\sim$1.5$\times$10$^{15}$cm$^{-2}$), and the mass of the molecule (4.83 $\times$ 10$^{-23}$ g for CH$_2$NH).
Using the desorption time scale t$_{\rm evap}$, the desorption rate R$_{\rm des}$(i) (cm$^{-3}$ $s^{-1}$) is given by t$_{\rm evap}^{-1}$n$_{\rm dust}$(i).
The desorption temperature is defined as the temperature which makes R$_{\rm des}$(i) and R$_{\rm acc}$(i) equal.
Assuming the grain radius of 1$\times$10$^{-5}$~cm$^{-1}$, the physical cross section of 3.14$\times$10$^{-10}$~cm$^{-2}$ is used for the $\sigma_d$.
In \cite{Ruaud15}, the E$_D$ for CH$_2$NH is assumed to be the same with CH$_3$OH (5530~K).
Although the mass of CH$_3$OH and CH$_2$NH are similar, this assumption may overestimate the desorption temperature for CH$_2$NH, since OH group in CH$_3$OH can strongly interact with H$_2$O molecules in the interstellar ice.
Thus, if we use the desorption temperature of CH$_3$OH, we would be able to estimate the upper limit to the desorption temperature for CH$_2$NH.

We used our simulation results by model~A, described in the section 4.2, to estimate the number density of CH$_2$NH (n$_{\rm gas}$ and n$_{\rm dust}$) just before the warm-up of the core.
Under the molecular hydrogen number density of 1$\times$10$^7$~cm$^{-3}$, n$_{\rm gas}$ (CH$_2$NH) and n$_{\rm dust}$ (CH$_2$NH) are roughly 1$\times$10$^{-7}$~cm$^{-3}$ and 1$\times$10$^{-4}$~cm$^{-3}$.
With this gas density, the dust number density n$_d$ can be estimated to be 1$\times$10$^5$~cm$^{-3}$ assuming the gas-to-dust ratio of 0.01.
Then, the upper limit to the desorption temperature can be calculated to be 160~K, which is comparable to the dust temperature in W3(H$_2$O).
Since CH$_2$NH would be sufficiently desorbed even in W3(H$_2$O), we can conclude that temperature differences among hot core sources could not explain our observational results.

The third hypothesis is that the different CH$_2$NH abundances are related to the evolutionary phases of HI\hspace*{-1pt}I regions inside a hot core.
In the ISM, it is well-known that star formation starts in a cold pre-stellar core.
The gas temperature gradually rises when a protostar is formed inside the core, and finally frozen species on the dust are evaporated into the gas phase when the dust temperature gets high enough.
As we have already mentioned, some of our sources are known to have high gas temperatures.
Followed by that, CH$_2$NH will be gradually photo-dissociated due to strong UV field in the vicinity of an evolved HI\hspace*{-1pt}I region inside the core.
In this case, the CH$_2$NH-rich sources can be explained as less evolved HI\hspace*{-1pt}I regions than the other sources, as is shown below.

\begin{figure}
\includegraphics[scale=.6]{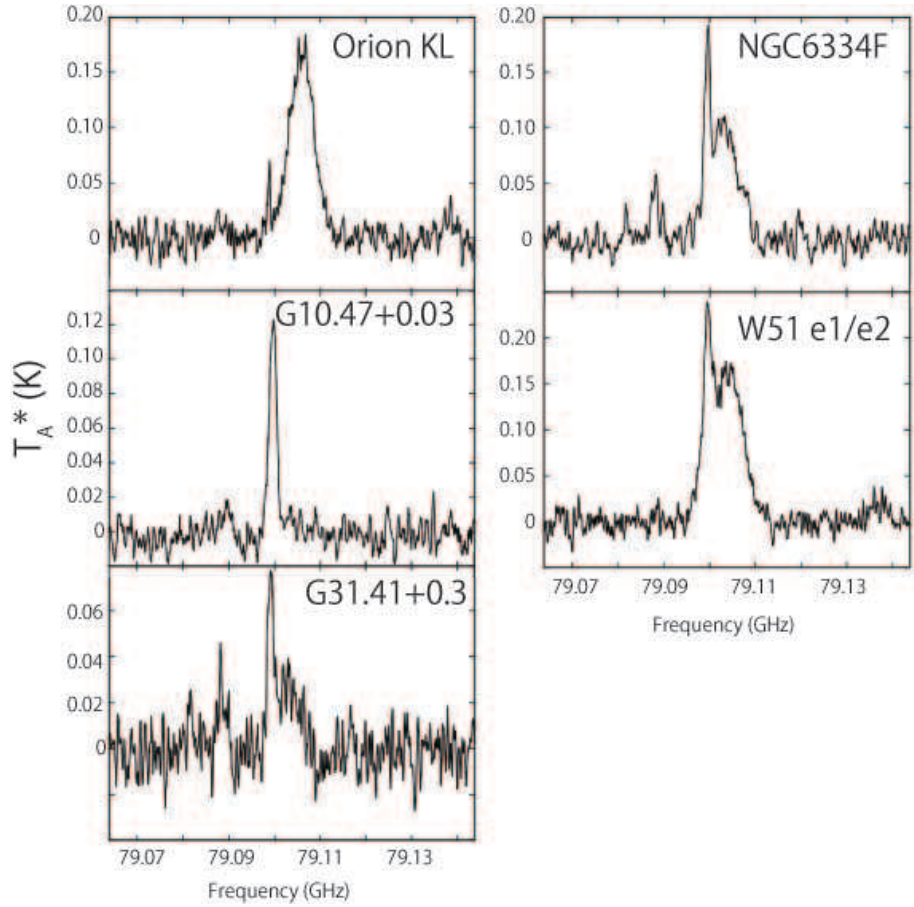}
\caption{
Observed recombination line ``H54$\beta$" is shown next to narrow CH$_3$CHO transition.
Its calculated frequency is 79.103866~GHz.
For the recombination line towards Orion KL, see the last part of section 4.1.
\label{fig:recombination_line}
}
\end{figure}

The hydrogen recombination lines, which are originated from HI\hspace*{-1pt}I regions, would be good tools to study the evolutionary state of an HI\hspace*{-1pt}I region.
We compared in Figure~\ref{fig:recombination_line} the strength of a recombination line, H54$\beta$, among five CH$_2$NH sources.
The observed line properties are summarized in Table~\ref{recombination_line}.
In G10.47+0.03 and G31.41+0.3, the observed H54$\beta$ line is very weak or below 3 sigma level.
For Orion~KL, it is well known that recombination lines are originated from the foreground HI\hspace*{-1pt}I region; \cite{Plambeck13} reported that recombination lines were not detected inside the Hot Core of Orion~KL.
Our results suggest that top three CH$_2$NH-rich sources would possess less evolved HI\hspace*{-1pt}I regions, while the H54$\beta$ line is prominent towards W51~e1/e2 and NGC6334F.
These sources are considered to have more evolved HI\hspace*{-1pt}I regions, supporting our scenario.
We also note that past observational results also agree with this scenario.
\cite{Cesaroni10} claimed that G10.47+0.03 contains two deeply embedded Hyper Compact HI\hspace*{-1pt}I regions in the positions of the hot core while no detectable HI\hspace*{-1pt}I region is present in G31.41+0.3.
Although it is difficult to compare directly with our observations, \cite{Jimenez-Serra11} detected H40$\alpha$, H34$\alpha$, and H31$\alpha$ towards Cep~A.
Cep~A would possess an evolved HI\hspace*{-1pt}I region inside and the dissociation process might result in the non-detection of CH$_2$NH.
Considering above, the difference of the evolutionary phase would be the most plausible reason to explain the different CH$_2$NH abundances.

\subsection{Astrochemical Modeling}
It is difficult to discuss actual formation processes of CH$_2$NH only from our observations.
In this section, we present the results of a gas-grain chemical model and discuss the chemistry of CH$_2$NH.

\subsubsection{The Nautilus Chemical Model}
To simulate the abundance of CH$_2$NH in hot cores, the NAUTILUS gas-grain chemical model \citep{Semenov10, Reboussin14} has been used with the similar physical model used in \cite{Garrod13} where a collapse phase is followed by a static warm-up phase.
This chemical model computes the time evolution of species in the gas-phase and at the surface of dust grains.
The gas-phase chemistry is described by the public network kida.uva.2014 \citep{Wakelam15}. 
During collisions with grains, species from the gas-phase can be physisorbed on the surfaces. They can then diffuse, react and be evaporated through thermal and non-thermal processes.
All details about the model can be found in \cite{Ruaud15}.
For our simulations, we have assumed that species are initially in the atomic form, except for H$_2$, with abundances from \cite{Garrod13}.
The initial abundance is listed in Table~\ref{initial_abundance}.

\subsubsection{The Physical Model}
The physical conditions used as input to the chemical model are based on the "fast warm-up model" from \cite{Garrod13}.
We have run four different models (A, B, C and D) with different peak density and/or temperature as listed in Table~\ref{models}. 
The peak temperatures and densities were selected considering the recent observations towards hot cores \citep[e.g.,][]{Hernandez14}.
We will see how CH$_2$NH formation processes depend on densities through the comparison of models~A, B, and C, and on temperatures through the comparison of models A and D.
The gas densities (n$_\mathrm{H_2}$) were initially set to be 3$\times$10$^3$ cm$^{-3}$, which gradually increased up to their peak densities along with the free-fall equation during collapsing phase.
While gas temperatures were fixed at 10~K in the collapsing phase, dust temperatures decreased from 16~K to 8~K depending on the visual extinction, which were calculated from the gas density of the cloud.
Once the peak densities were achieved, we fixed the densities, and the gas and dust temperatures were raised from $\sim$10~K to their peak temperatures in 7$\times$10$^4$ years.

\subsubsection{Gas and Ice Phase Formation of CH$_2$NH}
As a result of our model, CH$_2$NH is formed both in the gas phase and on the dust grain surface.
We summarize the main chemical reactions to form CH$_2$NH on the surfaces in Table~\ref{dust_reactions}. 
The dust surface reaction rates are proportional to both the probability of the reaction $\kappa$ and the diffusion rates of reactant species \citep{Hasegawa92}.
$\kappa$ is proportional to exp(-E$_A^{1/2}$), where E$_A$ represents the value of the activation barrier.
The diffusion rates depend on the exponential of the dust temperature.
For activation barriers of hydrogenation processes of HCN and CH$_2$NH, the theoretically predicted values by \cite{Woon02} were employed.

The main reactions in the gas-phase to form CH$_2$NH are summarized in Table~\ref{gas_reactions}.
CH$_2$NH is formed both by dissociative recombination processes and neutral-neutral reactions.
The rate coefficients of gas phase reactions are given by the Arrhenius equation below, using the coefficients $\alpha$, $\beta$, and $\gamma$:
\begin{equation}
k(T)=\alpha (T/300)^\beta e^{-\gamma/T},
\end{equation}
where T represents the gas temperature.
As shown in Table~\ref{gas_reactions}, the dissociative recombination processes have larger reaction rates.

\subsubsection{Modeling Results and Discussions}
\begin{figure}
\includegraphics[scale=.6]{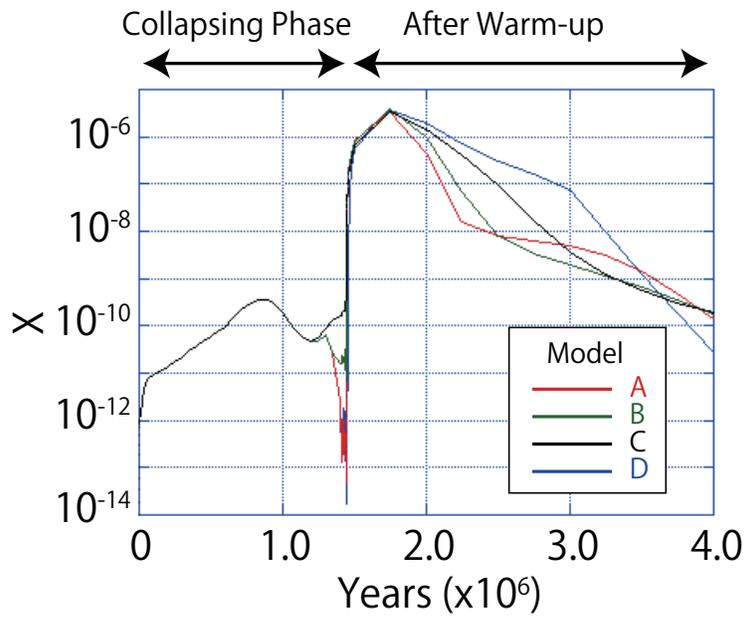}
\caption{
Simulated gas phase CH$_2$NH abundances compared to total hydrogen atoms are shown as ``X".
The red, blue, green, and black lines, respectively, represent the simulation results in models A, B, C and D.
\label{fig:simulation_results}
}
\end{figure}

The simulated gas phase CH$_2$NH abundances (relative to the total proton density; hereafter represented by a simbol X) for the four models A, B, C, and D are presented in Figure~\ref{fig:simulation_results} with different colors.
The origin in the X axis corresponds to the time of start the collapse of the gas.
We found that the general behavior of X[CH$_2$NH] along with the cloud evolution does not depend on the model.
X[CH$_2$NH] are almost the same in each model in the collapsing phase ($<$ 1.4$\times$10$^6$ years).
X[CH$_2$NH] decreases just before the warm-up phase ($\sim$ 1.4$\times$10$^6$ years), since gas phase species are accreted on the dust surface due to high gas density and low temperature environment.
After the warm-up phase, X[CH$_2$NH] increases dramatically.
It would be important to discuss the detailed CH$_2$NH formation process in this phase for understanding the dominant formation path.
Following that, X[CH$_2$NH] decreases via gas phase reactions with ions and/or radicals.

\begin{figure}
\includegraphics[scale=.6]{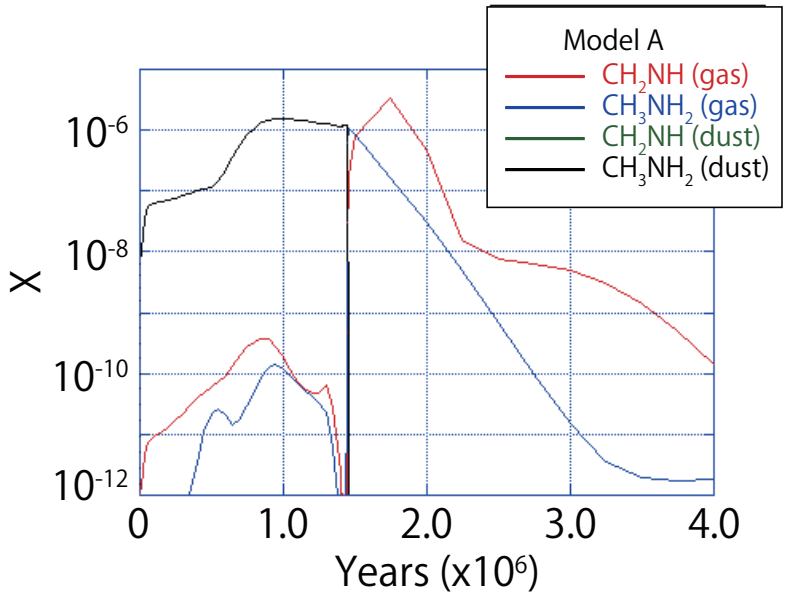}
\caption{
Species abundances (with respect to the total proton density) computed by model A as a function of time.
The red, blue, green, and black lines, respectively, correspond to the abundance of CH$_2$NH (ice), CH$_3$NH$_2$ (ice), CH$_2$NH (gas phase), and CH$_3$NH$_2$(gas phase).
The abundance of CH$_2$NH in the ice is lower than 10$^{-12}$ at any time.
\label{fig:simulation_gas_or_dust}
}
\end{figure}

Which chemical precesses are the most dominant formation process for CH$_2$NH?
First of all, in Figure~\ref{fig:simulation_gas_or_dust}, we compared X[CH$_2$NH] and X[CH$_3$NH$_2$] both in the gas phase and dust grain surface using model~A.
We can easily recognize that dust surface abundance before the warm-up phase ($\sim$1.4$\times$10$^{6}$ years) and gas phase abundance of CH$_3$NH$_2$ after warm-up are almost the same.
CH$_3$NH$_2$ would be mainly formed on the dust surface before the birth of a star, and evaporate to the gas phase. 
On the other hand, the dust surface X[CH$_2$NH] is very low ($<$ 10$^{-12}$) compared to the gas phase X[CH$_2$NH], since dust surface CH$_2$NH is efficiently hydrogenated to CH$_3$NH$_2$ during the collapsing phase.
We also confirmed the same trend in models~B, C, and D.
Thus, gas phase reactions would be dominant formation process to explain gas phase X[CH$_2$NH] in the protostellar stage.

\begin{figure}
\includegraphics[scale=.6]{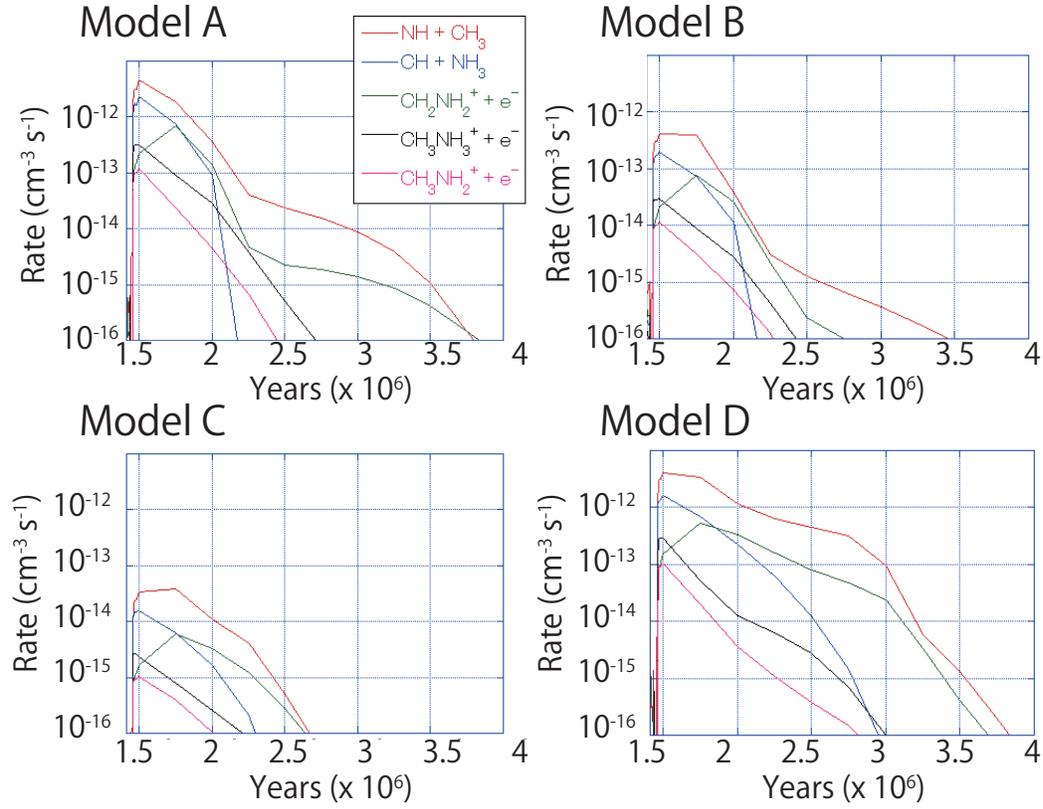}
\caption{
Formation rates of CH$_2$NH rates as a function of time after warm-up for each model.
The red, blue, green, black and pink lines, respectively, correspond to the CH$_2$NH formation rates by NH + CH$_3$, CH + NH$_3$, CH$_2$NH$_2^+$ + e$^-$, CH$_3$NH$_3^+$ + e$^-$, and CH$_3$NH$_2^+$ + e$^-$.
\label{fig:simulation_which_gas}
}
\end{figure}

To discuss the most dominant gas phase reactions in Table~\ref{gas_reactions}, we compared their formation rates k$_{rate}$(T, t) (cm$^{-3}$ s$^{-1}$).
Formation rates are presented by  k$_{rate}$(T, t) = k(T) $\times$ n$_1$(t) $\times$ n$_2$(t), where k(T) is the temperature-dependent rate constant (cm$^{3}$ s$^{-1}$) calculated with the Arrhenius equation, and n$_1$(t) and n$_2$(t) are the number densities (cm$^{-3}$) of reactant species in each time step.
In Figures~\ref{fig:simulation_which_gas} (A)-(D), time dependent behaviors of k$_{rate}$(T, t) are shown for models~A, B, C, and D, after the collapsing phase (t$>$1.4$\times$10$^{6}$~years).
In these figures, the formation rates of NH + CH$_3$, CH + NH$_3$, CH$_2$NH$_2^+$ + e$^-$, CH$_3$NH$_3^+$ + e$^-$, and CH$_3$NH$_2^+$ + e$^-$ are compared.
As we can see through the comparison of models~A and C, formation rates tend to be lower under low density environment, due to low number densities of reactant species.
The difference between model~A and model~D is originated from the different rate coefficients determined by the gas temperatures. 
However, we confirmed that NH + CH$_3$ reactions would be the most dominant formation process in all the models.

\begin{figure}
\includegraphics[scale=.45]{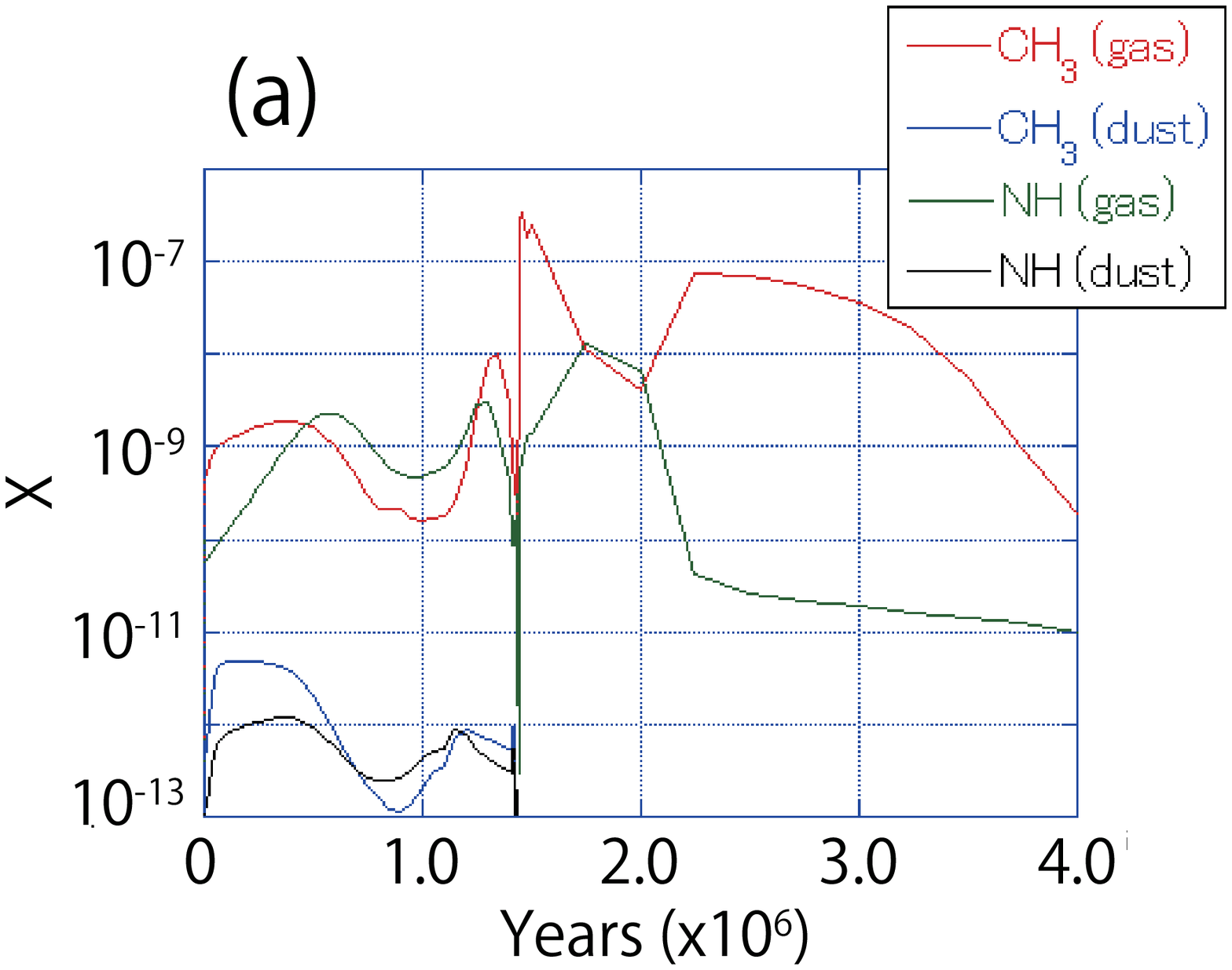}
\includegraphics[scale=.45]{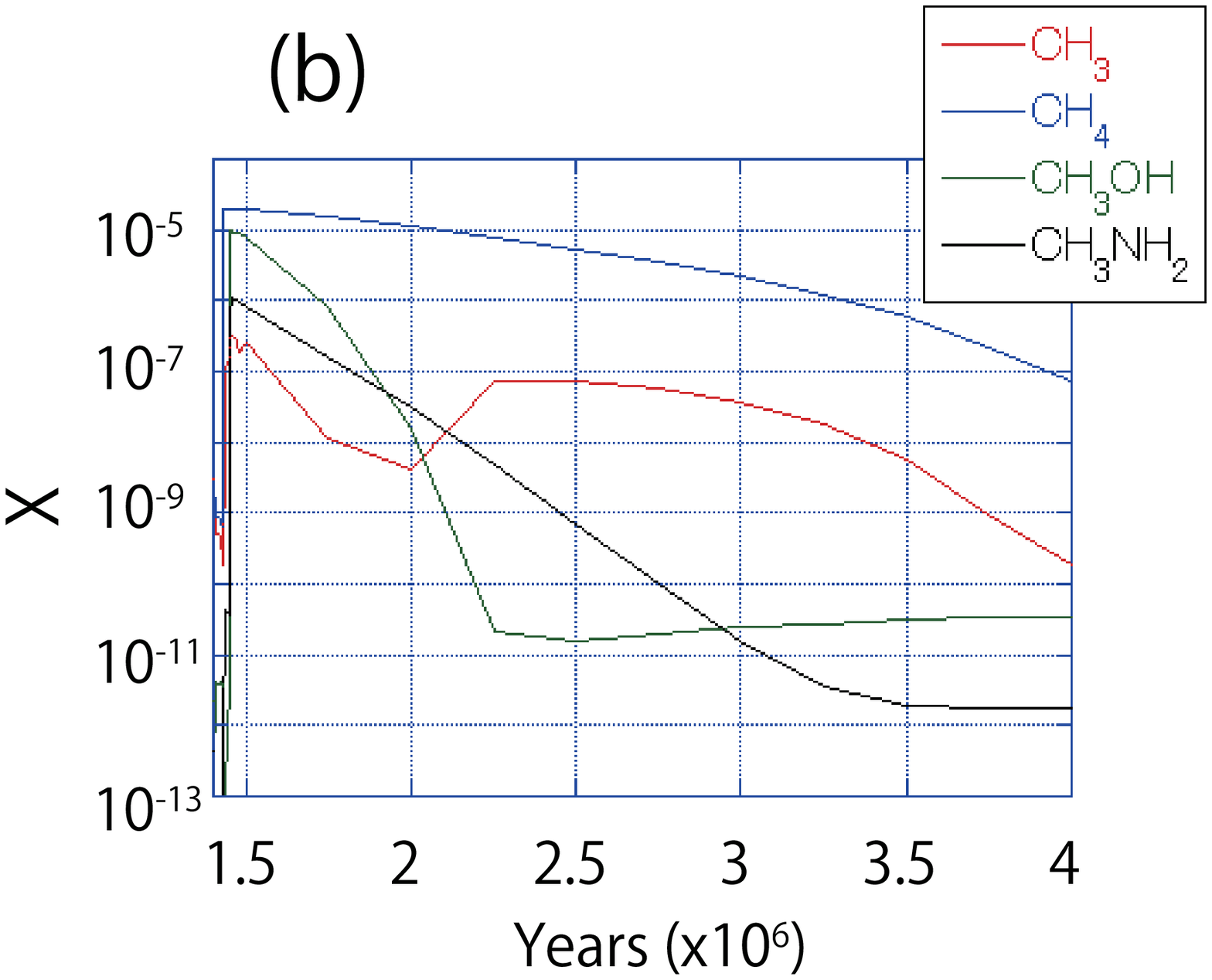}
\includegraphics[scale=.45]{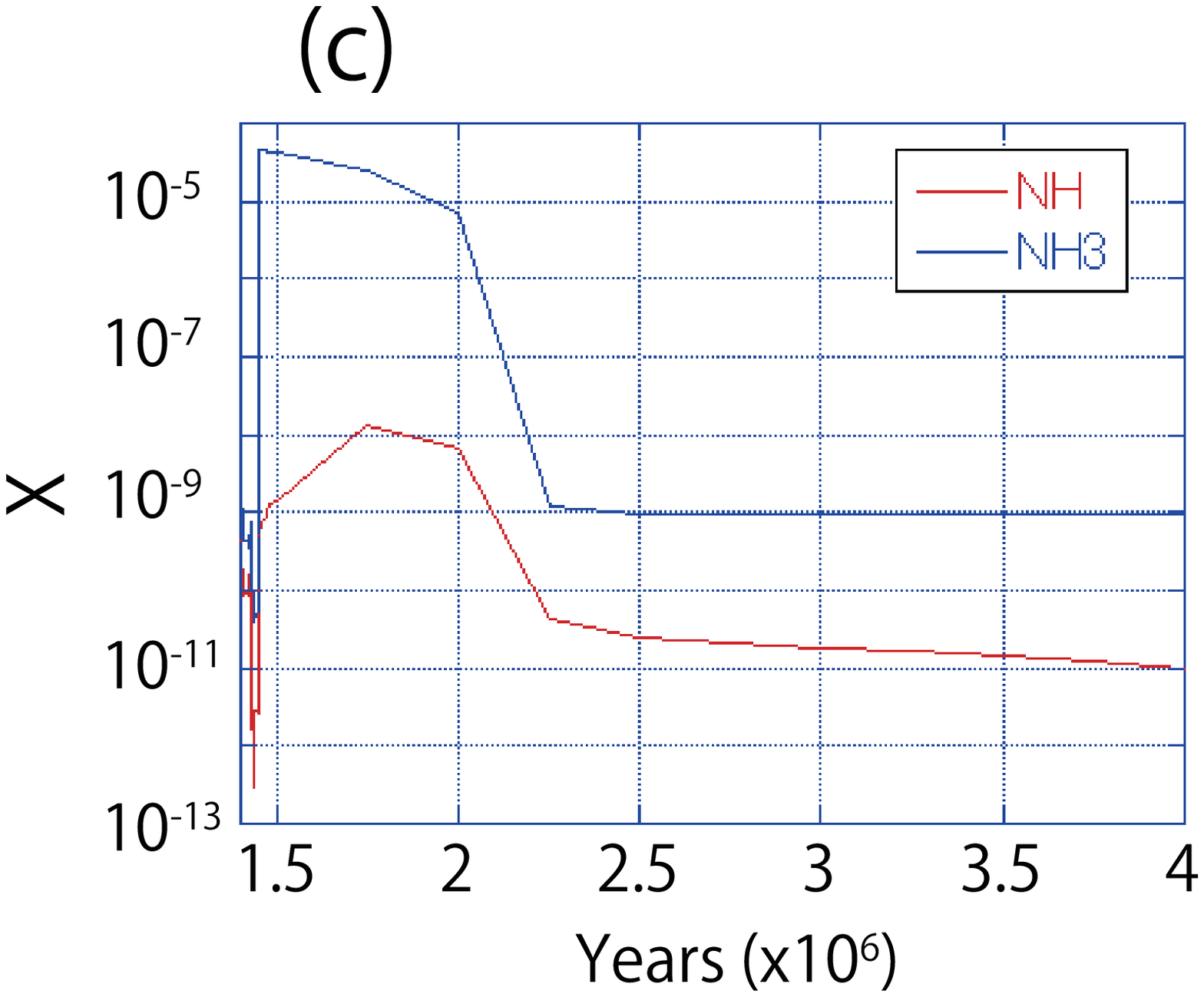}
\caption{
Using model~A, we show simulated abundances of chemical species related to CH$_3$ and NH after warm-up phase.
``X" represent the abundances compared to the total hydrogen atoms.
In (a), red, blue, green, and black lines, respectively, represent CH$_3$ abundance in gas phase and on dust grain surface, and NH abundances in gas phase and on dust grain surface.
In (b), red, blue, green, and black lines, respectively, represent gas phase abundances of CH$_3$, CH$_4$, CH$_3$OH, and CH$_3$NH$_2$ after warm-up.
In (c), red and blue lines respectively show the gas phase abundances of NH and NH$_3$ after warm-up.
\label{fig:simulation_radical}
}
\end{figure}

Finally, we will mention the origin of these radicals.
The dust surface abundances of radicals are very low during the collapsing phase, since unstable species are quickly hydrogenated.
In Figure~\ref{fig:simulation_radical} (A), we show the X[CH$_3$] in the gas phase and on the dust surface with red and blue lines, and the X[NH] in the gas phase and on the dust surface with green and black lines.
The dust surface X[CH$_3$] and X[NH] are both nearly 10$^{-12}$ before the warm-up phase ($\sim$1.4$\times$10$^{6}$ years).
Since X[CH$_3$] and X[NH] in the gas phase are much higher than 10$^{-12}$ after the warm-up phase, the gas phase reactions would be essential to explain the abundant X[CH$_3$] and X[NH]. 
In the dataset of KIDA, radicals are formed via the destructions of complex species.
We show the gas phase abundances of some species are related to the formations of CH$_3$ and NH using model~A, in Figures~\ref{fig:simulation_radical} (B) and (C).
CH$_3$ radical is originated from the destruction of larger saturated molecules like CH$_4$, CH$_3$OH, and CH$_3$NH$_2$ by atoms and/or secondary UV radiation  (Figure~\ref{fig:simulation_radical} (B)).
NH radical is formed via the destruction of NH$_3$  (Figure~\ref{fig:simulation_radical} (C)).

We summarize the CH$_2$NH formation process here.
During the cold collapsing phase, saturated complex molecules like NH$_3$, CH$_4$, CH$_3$OH, and CH$_3$NH$_2$ are formed on the dust surface.
We note that dust surface CH$_2$NH abundance is quite low in the collapsing phase.
During the warm-up phase, these molecules are evaporated to the gas, and NH and CH$_3$ radicals are produced through the gas phase reactions using these evaporated molecules.
Finally, newly formed radicals form CH$_2$NH.

\section{Conclusion}
The main results of this paper can be summarized as follows:

\begin{enumerate}
\item We conducted the survey observations of CH$_2$NH towards high-mass and low-mass star-forming regions; CH$_2$NH were detected in eight high-mass star-forming regions. 
Among them, the first detections were made in four objects, G10.47+0.03, G31.41+0.3, DR21(OH), and NGC6334F.
However, CH$_2$NH was not detected towards low-mass star-forming regions.

\item We calculated the column densities and fractional abundances or their upper limits in our sources.
The fractional abundances range from $\sim$10$^{-9}$ to $\sim$10$^{-8}$.
We also found that Orion KL and G10.47+0.03 have higher CH$_2$NH abundances than other sources by almost one order of magnitude.

\item CH$_2$NH-rich sources, Orion KL and G10.47+0.03, have high temperature regions, while HI\hspace*{-1pt}I regions are less evolved than other sources.
CH$_2$NH would be much less likely photo dissociated by UV photons in such an environment.

\item From the results of our chemical model, it appears that CH$_2$NH in the hot sources seems to have been formed in the gas-phase rather than evaporated from the ices.
First, saturated molecules are evaporated from the dust surface during the warm-up phase of the cloud.
Second, such species are destructed in the gas phase to form radicals.
Following that, the reaction between CH$_3$ and NH radicals form CH$_2$NH.
\end{enumerate}

\acknowledgments
We are grateful all the staff members of the Nobeyama Radio Observatory, the National Astronomical 
Observatory of Japan (NAOJ), and the Arizona Radio Observatory for their support throughout our observations.
We thank to Dr.Hideko Nomura, and taffs in Bordeaux University for fruitful discussions.
We thank to Drs. Robbin Garrod and Eric Herbst for providing us with their copies of chemical reaction data file.
We also thank the anonymous referee for the critical reading, which has improved this paper so much.
A part of the data analysis was made at the Astronomy Data Center, NAOJ.
This research has made use of NASA's Astrophysics Data System.
This study was supported by the Astrobiology Program of National Institutes of Natural Sciences (NINS) and by the JSPS Kakenhi Grant  Numbers 15H03646 and 14J03618.
Valentine Wakelam and Liton Majumdar acknowledge the European Research Council (3DICE, grant agreement 336474) and the French CNRS/INSU programme PCMI for fundings.





\appendix

\clearpage

\clearpage

\clearpage

\clearpage

\clearpage

\clearpage

\clearpage

\clearpage

\clearpage

\clearpage

\clearpage

\clearpage

\clearpage

\begin{deluxetable}{cccc}
\tabletypesize{\scriptsize}
\tablecaption{Detected CH$_2$NH Lines in the Past Survey}
\tablewidth{0pt}
\tablehead{
\colhead{Source }
& \colhead{\shortstack {transitions}} 
& \colhead{\shortstack {$\nu$ \\ (MHz)} } 
& \colhead{\shortstack {reference}} 
}
\startdata
Sagittarius B2$^*$ &  & & (1) \\
Orion KL & 2$_{ 1 1}$-2$_{ 0 2}$ & 172267.113 & (2) \\
& 1$_{ 1 0}$-1$_{ 1 1}$ & 225554.692 & (2) \\
& 4$_{ 1 4}$-3$_{ 1 3}$ & 245125.974 & (2) \\
& 7$_{ 1 6}$-7$_{ 0 7}$ & 250161.865 &  (2) \\
& 6$_{ 0 6}$-5$_{ 1 5}$ & 251421.379 & (2) \\
W51 e1/e2 & 1$_{ 1 1}$-0$_{ 0 0}$ & 225554.692 & (2) \\
& 4$_{ 1 4}$-3$_{ 1 3}$ & 245125.974 & (2) \\
& 6$_{ 0 6}$-5$_{ 1 5}$ & 251421.379 & (2) \\
& 4$_{ 2 3}$-3$_{ 2 2}$ & 255840.431 & (2) \\
G34.3+0.2 & 4$_{ 1 4}$-3$_{ 1 3}$ & 245125.974 & (2) \\
G19.61-0.23 & 3$_{ 1 3}$-2$_{ 0 2}$ & 340354.315 & (3) \\
\\
\enddata
\tablecomments{
$^*$27 transitions were reported by \cite{Halfen13} towards Sagittarius B2.
 References. (1) \cite{Halfen13} and references therein (2) \cite{Dickens97} (3) \cite{Qin10}
}
\label{past_CH2NHlines}
\end{deluxetable}

\begin{deluxetable}{ccccc}
\tabletypesize{\scriptsize}
\tablecaption{Observed CH$_2$NH Lines}
\tablewidth{0pt}
\tablehead{
\colhead{Transition }
& \colhead{\shortstack {$\nu$ \\ (MHz)} } 
& \colhead{\shortstack {E$_u$ \\ (K)} } 
& \colhead{\shortstack {$S\mu^2$ \\ (D$^2$)}} 
& \colhead{Telescope }
}
\startdata
4$_{ 0 4}$-3$_{ 1 3}$ & 105794.062 &  30.6 & 3.93 & NRO \\
1$_{ 1 1}$-0$_{ 0 0}$ & 225554.609 & 10.8 & 2.34 & SMT \\
4$_{ 1 4}$-3$_{ 1 3}$ & 245125.866 & 37.3 & 6.58 & SMT \\
4$_{ 2 3}$-3$_{ 2 2}$ & 255840.311 & 62.2 &  5.27 & SMT \\
\\
\enddata
\label{CH2NHlines-summary}
\end{deluxetable}

\begin{deluxetable}{ccccccccc}
\tabletypesize{\scriptsize}
\tablecaption{List of Observed Sources}
\tablewidth{0pt}
\tablehead{
\colhead{Source }
& \colhead{\shortstack {$\alpha$(J2000) \\  $^{\rm h}$ $^{\rm m}$ $^{\rm s}$}  }
& \colhead{\shortstack {$\delta$(J2000)  \\  $\degr$  '  "} } 
& \colhead{\shortstack {V$_{\rm LSR}$ \\ (km~s~$^{-1}$)}} 
& \colhead{\shortstack {distance \\ (kpc)}}
& \colhead{Masers} 
& \colhead{{\shortstack {Mass \\ High/Low}}} 
& \colhead{Detection} 
& \colhead{reference} 
}
\startdata
W3(H$_2$O)&02 27 04.6&+61 52 25&-47&2.0&M&H&no&1, 6\\
NGC1333 IRAS4B&03 29 11.99&+31 13 08.9&6.8&0.2&&L&no&3, 7\\
Orion KL&05 35 14.5&-05 22 30.6&6&0.4&M&H&confirm&1, 8\\
NGC2264 MMS3&06 41 12.3&+09 29 11.9&8&0.7&&H&no&4, 9\\
IRAS16293-2422&16 32 22.63&-24 28 31.8&2.7&0.12&&L&no&5, 10\\
NGC6334F&17 26 42.8&-36 09 17.0&-7&1.3&M&H&new&1, 6\\
G10.47+0.03&18 08 38.13&-19 51 49.4&68&8.5&M&H&new&1, 6\\
G19.61-0.23&18 27 38.0&-11 56 42&41.9&4.0&M&H&confirm&1, 11\\
G31.41+0.3&18 47 34.6&-01 12 43&97&7.9&M&H&new&1, 12\\
G34.3+0.2&18 53 18.54&+01 14 57.0&58&1.6&M&H&confirm&1, 6\\
W51 e1/e2&19 23 43.77&+14 30 25.9&58&5.4&M&H&confirm&1, 6\\
G75.78+0.34&20 21 44.1&+37 26 40&-0.1&3.8&M&H&no&2, 6\\
DR21 (OH)&20 39 01.1&+42 22 50.2&-3&1.5&M&H&new&1, 6\\
Cep A&22 56 17.9&+62 01 49&-10.5&0.7&M&H&no&2, 6\\
\enddata
\tablecomments{
``M" represents CH$_3$OH maser sources reported in \cite{Minier02}. ``H" and ``L" respectively shows high-mass and low-mass star-forming regions. In the column of detection, ``no",``confirm", and ``new", respectively, represent non-detection sources, confirmations of the past CH$_2$NH surveys, and new CH$_2$NH detections.
 References. (1) \cite{Ikeda01} (2) \cite{Minier02} (3) \cite{Sakai06} (4) \cite{Sakai07} (5)\cite{Cauzax03} (6)\cite{Reid14} (7)\cite{Hirota08} (8)\cite{Menten07} (9) \cite{Sung97} (10)\cite{Loinard08} (11)\cite{Hofner96} (12)\cite{Churchwell90}
}
\label{sources}
\end{deluxetable}

\begin{deluxetable}{cccccc}
\tabletypesize{\scriptsize}
\tablecaption{Observed Line Parameters of CH$_2$NH 4$_{ 0, 4}$-3$_{ 1, 3}$ transition at 105.794062~GHz}
\tablewidth{0pt}
\tablehead{
\colhead{Source}
& \colhead{\shortstack {Obs. \\ freq \\ (GHz)}}  
& \colhead{\shortstack {T$_{\rm A}$* \\ (mK) }}
& \colhead{\shortstack {$\Delta v$ \\ (km~s~$^{-1}$) }}
& \colhead{\shortstack {V$_{\rm LSR}$ \\ (km~s~$^{-1}$)}} 
& \colhead{\shortstack {rms \\ (mK) }}
}
\startdata
DR21(OH)&105.79383&24 &5.4&-2.4&7 \\
G10.47+0.03&105.79429&99 &7.7&67.4&11 \\
G31.41+0.3&105.79404&98 &2.4&97.1&24 \\
G34.3+0.2&105.79384&42 &9.7&58.6&14 \\
NGC6334F&105.79420&47 &1.8&-7.4&19 \\
Orion KL&105.79352&108 &9.3&7.5&13 \\
W51~e1/e2&105.79406&54 &11.6&57.0&15 \\
G19.61-0.23&105.79497&31 &8.5&39.3&12 \\
\enddata
\label{CH2NHline-obs}
\end{deluxetable}
\clearpage

\begin{deluxetable}{cccccc}
\tabletypesize{\scriptsize}
\tablecaption{CH$_2$NH Abundances for Compact Sources}
\tablewidth{0pt}
\tablehead{
\colhead{Source}
& \colhead{\shortstack {Size \\ (components)}}
& \colhead{\shortstack {N[H$_2$] \\ (10$^{24}$cm$^{-2}$)}}
& \colhead{\shortstack {T$_{\mathrm{ex}}$ \\ (K)}} 
& \colhead{\shortstack {N[CH$_2$NH] \\ (10$^{16}$ cm$^{-2}$)}}
& \colhead{\shortstack {X[CH$_2$NH] \\  (10$^{-8}$)}}
}
\startdata
Orion KL&2" (3)&1.4$^{a}$&88 ($\pm$23)&4.6($\pm$0.7)&3.3($\pm$0.5)\\
G10.47+0.03&1.4"&6.7$^{b}$&84 ($\pm$57)&21($\pm$7.3)&3.1($\pm$1.1)\\
G31.41+0.3&1.7"&3.5$^{b}$&28 ($\pm$17)&3.1($\pm$3.0)&0.88($\pm$0.85)\\
W51~e1/e2&1.2" (2)&4.0$^{b}$&20 ($\pm$2)&1.1($\pm$0.63)&0.28($\pm$0.16)\\
NGC6334F&3.7"&1.4$^{b}$&[40]&0.3($\pm$0.2)&0.24($\pm$0.14)\\
G19.61-0.23&2.5" &6.7$^{c}$&[40]&0.94($\pm$0.65)&0.14($\pm$0.10)\\
\enddata
\tablecomments{
CH$_2$NH column densities ``N", or its upper limits and fractional abundances ``X" are shown.
The distributions of CH$_2$NH were assumed to be equal to dust continuum.
Since it is known that Orion~KL has three CH$_2$NH peaks and W51~e1/e2 has two continuum components, beam coupling factors were modified for these sources.
Excitation temperatures are fixed at 40~K shown with square brackets if only one transition was available, and the uncertainty associated in this assumption is assumed to be 50 percent of its central value considering wide range of excitation temperatures derived with rotation diagram method.
\newline
{\bf References.} (a)\cite{Hirota15} (b) \cite{Hernandez14} (c) \cite{Wu09}
}
\label{CH2NH-abundance-compact}
\end{deluxetable}
\clearpage

\begin{deluxetable}{cccccc}
\tabletypesize{\scriptsize}
\tablecaption{CH$_2$NH Abundances for 10" Sources}
\tablewidth{0pt}
\tablehead{
\colhead{Source}
& \colhead{\shortstack {N[H$_2$] \\ (10$^{23}$cm$^{-2}$)}}
& \colhead{\shortstack {T$_{\mathrm{ex}}$ \\ (K)}} 
& \colhead{\shortstack {N[CH$_2$NH] \\ (10$^{15}$ cm$^{-2}$)}}
& \colhead{\shortstack {X[CH$_2$NH] \\  (10$^{-8}$)}}
& \colhead{\shortstack {X$_{\mathrm{compact}}$/X$_{\mathrm{10sec}}$}}
}
\startdata
Orion KL&1.0$^{a}$&88 ($\pm$23)&5.5($\pm$0.8)&5.5($\pm$0.8)&0.6\\
G10.47+0.03&1.3$^{a}$&84 ($\pm$57)&4.7($\pm$1.6)&3.6($\pm$1.3)&0.9\\
G31.41+0.3&1.6$^{a}$&28 ($\pm$17)&0.85($\pm$0.82)&0.53($\pm$0.51)&1.7\\
NGC6334F&2.0$^{a}$&[40]&0.3($\pm$0.2)&0.16($\pm$0.12)&1.1\\
W51~e1/e2&3.6$^{a}$&20 ($\pm$2)&0.32($\pm$0.09)&0.09($\pm$0.03)&3.1\\
\hline
G34.3+0.2&3.0$^{a}$&[40]&0.7($\pm$0.4)&0.24($\pm$0.14)\\
DR21(OH)&2.0$^{a}$&[40]&0.27($\pm$0.21)&0.14($\pm$0.11)\\
W3(H2O)&1.0$^{a}$&[40]&$<$0.7&$<$0.7\\
G75.78+0.34&1.0$^{e}$&[40]&$<$0.6&$<$0.6\\
CepA&1.0$^{e}$&[40]&$<$0.5&$<$0.5\\
NGC2264 MMS3&5.7$^{b}$&[40]&$<$0.1&$<$0.02\\
NGC1333 IRAS4B&3.0$^{c}$&[40]&$<$0.1&$<$0.04\\
IRAS16293-2422&0.8$^{d}$&[40]&$<$0.1&$<$0.2\\
\enddata
\tablecomments{
For the sources where dust continuum data are not available, a source size of 10" was assumed to derive the fractional abundances from the spatially extended CO observations.
Although we have already obtained more reliable fractional abundance towards Orion KL, G10.47+0.03, G31.41+0.3, NGC6334F, and W51~e1/e2 in Table~\ref{CH2NH-abundance-compact}, we also calculated them assuming 10" sources to know how fractional abundance can change depending on the source sizes.
As the results, the differences of the fractional abundances in two different source sizes were within a factor of 3.
The upper limits of CH$_2$NH are shown for non-detection sources.
\newline
{\bf References.} (a)\cite{Ikeda01} (b) \cite{Peretto06} (c) \cite{Jorgensen02} (d) \cite{Ceccarelli00} (e)Assumed value
}
\label{CH2NH-abundance-10sec}
\end{deluxetable}
\clearpage

\begin{deluxetable}{ccccc}
\tabletypesize{\scriptsize}
\tablecaption{Observed Parameters of H54$\beta$}
\tablewidth{0pt}
\tablehead{
\colhead{Source} 
& \colhead{\shortstack {Tb \\ (mK)}}
& \colhead{\shortstack {$\Delta v$\\ (kms$^{-1}$)}}
& \colhead{\shortstack {W \\ (mK$\cdot$kms$^{-1}$)}}
& \colhead{\shortstack {X[CH$_2$NH] \\ $\times$ 10$^{-8}$}}
}
\startdata
Orion~KL&Not detected$^*$ &&&3.3($\pm$0.5)\\
G10.47+0.3&$<$24&[30]&$<$72&3.1($\pm$1.1)\\
G31.41+0.03&131&23.6&1018&0.89($\pm$0.85)\\
NGC6334F&383&29.7&3762&0.24($\pm$0.17)\\
W51~e1/e2&531&30.3&5306&0.28($\pm$0.16)\\
\enddata
\tablecomments{
Observed recombination line parameters, brightness temperatures Tb (mK), line width $\Delta v$ (kms$^{-1}$), and integrated intensities W (mK$\cdot$kms$^{-1}$) are shown.
The upper limit was calculated for G10.47+0.03, with 3 $\sigma$ noise level and line width of 30 kms$^{-1}$.  
We also showed the fractional abundances of CH$_2$NH from Table~\ref{CH2NH-abundance-compact}.
$^*$Reference: \cite{Plambeck13}.
}
\label{recombination_line}
\end{deluxetable}

\begin{deluxetable}{cccc}
\tablecaption{Initial Elemental Abundances Compared to Total Hydrogen Atoms}
\tablewidth{0pt}
\tablehead{
\colhead{Element} & \colhead{Abundance} & \colhead{Element} & \colhead{Abundance}
}
\startdata
H$_2$ & 0.499 & Na & 2.0(-8) \\
H & 2.0(-3) & Mg & 7.00(-9) \\
He & 9.00(-2) & Si & 8.0(-9) \\
C & 1.40(-4) & P & 3.0(-9) \\
N & 7.50(-5) & Cl & 4.00(-9)\\
O & 3.2(-4) & Fe & 3.0(-9) \\
S & 8.0(-8) & &
\enddata
\tablecomments{Elemental abundance used in our chemical reaction model. 
This table is referred from \cite{Garrod13}.
}
\label{initial_abundance}
\end{deluxetable}

\begin{deluxetable}{ccc}
\tablecaption{Peak Temperatures and Densities for the Models}
\tablewidth{0pt}
\tablehead{
\colhead{Model}
& \colhead{\shortstack {Peak Density \\ (cm$^{-3}$)}}
& \colhead{\shortstack {Peak Temperature \\ (K)}}
}
\startdata
A & 1$\times$10$^{7}$ & 400 \\ 
B & 1$\times$10$^{6}$ & 400 \\
C & 1$\times$10$^{5}$ & 400 \\
D & 1$\times$10$^{7}$ & 200 \\  
\enddata
\label{models}
\end{deluxetable}

\begin{deluxetable}{ll}
\tablecaption{Dust Surface Reactions Related to CH$_2$NH }
\tablewidth{0pt}
\tablehead{
\colhead{Reaction}
& \colhead{\shortstack {E$_A$ \\ (K)}}
}
\startdata
N + CH$_3$ $\longrightarrow$ CH$_2$NH& E$_A$= 0~K\\
NH + CH$_2$ $\longrightarrow$ CH$_2$NH& E$_A$= 0~K\\
NH$_2$ + CH $\longrightarrow$ CH$_2$NH& E$_A$= 0~K\\
HCN + H $\longrightarrow$ H$_2$CN& E$_A$= 3647~K \\
HCN + H $\longrightarrow$ HCNH& E$_A$= 6440~K \\
H$_2$CN + H $\longrightarrow$ CH$_2$NH& E$_A$= 0~K \\
HCNH + H $\longrightarrow$ CH$_2$NH& E$_A$= 0~K \\
CH$_2$NH + H $\longrightarrow$ CH$_3$NH& E$_A$= 2134~K \\
CH$_2$NH + H $\longrightarrow$ CH$_2$NH$_2$& E$_A$= 3170~K \\
CH$_3$NH + H $\longrightarrow$ CH$_3$NH$_2$& E$_A$= 0~K \\
CH$_2$NH$_2$ + H $\longrightarrow$ CH$_3$NH$_2$& E$_A$= 0~K \\
\enddata
\tablecomments{
The dust surface reactions related to CH$_2$NH are shown.
E$_A$ represents the value of the activation barrier.
Since radical species are so reactive, radical-radical reactions would have no activation barriers.
The activation barriers for HCN and CH$_2$NH were cited from the theoretical study by \cite{Woon02}.
}
\label{dust_reactions}
\end{deluxetable}

\begin{deluxetable}{llll}
\tablecaption{Gas Phase Reactions Related to CH$_2$NH }
\tablewidth{0pt}
\tablehead{
\colhead{Reaction}
& \colhead{$\alpha$}
& \colhead{$\beta$}
& \colhead{$\gamma$}
}
\startdata
NH + CH$_3$ $\longrightarrow$ CH$_2$NH + H & 1.3$\times$10$^{-10}$ & 1.7$\times$10$^{-1}$ & 0 \\
CH + NH$_3$ $\longrightarrow$ CH$_2$NH + H & 1.52$\times$10$^{-10}$ & -5$\times$10$^{-2}$ & 0 \\
CH$_2$NH$_2^{+}$ + e$^{-}$ $\longrightarrow$ CH$_2$NH + H & 1.5$\times$10$^{-7}$ & -5$\times$10$^{-1}$ & 0 \\
CH$_3$NH$_3^{+}$ + e$^{-}$ $\longrightarrow$ CH$_2$NH + H$_2$ + H & 1.5$\times$10$^{-7}$ & -5$\times$10$^{-1}$ & 0 \\
CH$_3$NH$_2^{+}$ + e$^{-}$ $\longrightarrow$ CH$_2$NH + H$_2$ & 1.5$\times$10$^{-7}$ & -5$\times$10$^{-1}$ & 0 \\
\enddata
\tablecomments{
$\alpha$, $\beta$, and $\gamma$ represent the coefficients for the Arrhenius equation.
}
\label{gas_reactions}
\end{deluxetable}




\end{document}